\documentclass[12pt]{article}%
\usepackage{amsfonts}
\usepackage{amsmath}
\usepackage{amssymb}
\usepackage{graphicx}%
\providecommand{\U}[1]{\protect\rule{.1in}{.1in}}
\begin{document}

\title{Signed Sequential Rank Shiryaev-Roberts Schemes}
\author{C. van Zyl\\Centre for Business Mathematics and Informatics \\North-West University, Potchefstroom 2520\\F. Lombard\\Department of Statistics\\University of Johannesburg, Johannesburg 2000}
\date{}
\maketitle

\begin{abstract}
We develop Shiryaev-Roberts schemes based on signed sequential ranks to detect
a persistent change in location of a continuous symmetric distribution with
known median. The in-control properties of these schemes are distribution
free, hence they do not require a parametric specification of an underlying
density function or the existence of any moments. Tables of control limits are
provided. The out-of-control average run length properties of the schemes are
gauged via theory-based calculations and Monte Carlo simulation. Comparisons
are made with two existing distribution-free schemes. We conclude that the 
newly proposed scheme has much to recommend its use in practice. Implementation of the
methodology is illustrated in an application to a data set from an
industrial environment. \bigskip


\newpage

\end{abstract}

\section{Introduction}

Statistical process control schemes, including CUSUMs and Shiryaev-Roberts
schemes, are designed to signal a persistent change in some characteristic of
a process as soon as possible after its onset. The change manifests itself as
a sustained change in the distribution along a sequence of independent and
identically distributed observations $X_{1},X_{2},\dots$ which occurs at an
index $\tau$, known as the the change point. Thus $X_{1},X_{2},\ldots,X_{\tau}%
$, the \textquotedblleft in-control\textquotedblright\ observations, have a
common density function $f$ \ while $X_{\tau+1},X_{\tau+2},\ldots$, the
\textquotedblleft out-of-control\textquotedblright\ observations, have common
density $g\neq f$. Perhaps the most widely known CUSUM is the one introduced
by Page (1954) which aims to detect an increase in the mean from $\delta=0$ to
a value $\delta>0$ in a normal distribution with known variance $\sigma^{2}$.
The CUSUM sequence $C_{i},$ $i\geq0,$ is defined by setting $C_{0}=0$ and
applying the recursion
\begin{equation}
C_{i}=\max\left\{  0,C_{i-1}+\frac{X_{i}}{\sigma}-\zeta\right\}  ,\ i\geq1.
\label{Cusum recursion}%
\end{equation}
An alarm is raised as soon as $C_{i}$ exceeds a control limit $h>0$,
indicating that the mean has possibly increased and that an out-of-control
situation prevails. Here, $\zeta$ is a positive reference value or drift
allowance. This is often set equal to $\delta_{1}/2$ where $\delta_{1}$, the
target shift, is the smallest change (drift) in the mean of $X/\sigma$ that is
judged to be of practical import. Alternatively, $\zeta$ is sometimes chosen
with a view to producing a specified form of out-of-control behaviour. The
CUSUM behaves like a random walk with a reflecting barrier at $0$ and is sure
to produce an alarm somewhere along the sequence even if no change ever
occurs. In the latter case, we have a false alarm. Because false alarms are
inevitable, the control limit $h$ is chosen so that a predetermined expected
run length, known as the in-control average run length (ICARL), is guaranteed
if no change ever occurs.

Girshick and Rubin (1952) introduced a control scheme based on a recursion
that can be expressed as
\begin{equation}
D_{i}=log(1+\exp(D_{i-1}))+2\zeta\left(  \frac{X_{i}}{\sigma}-\zeta\right)
,\ i\geq1 \label{S-R recursion}%
\end{equation}
with $D_{0}=-\infty$ . A continuous time version of this scheme was developed
by Shiryaev (1963)\emph{\ }with the objective\emph{ }of detecting the onset of
a drift in a Brownian motion. Roberts (1966) compared the performance of the
CUSUM and some other schemes with the scheme (\ref{S-R recursion}) and found
the latter to have merit. Schemes based on the recursion (\ref{S-R recursion})
have subsequently become known in the literature as Shiryaev-Roberts (S-R) schemes.

The assumption of normality and of a known variance are enduring problems in
the application of these schemes to observed data. Versions that are
distribution free under a broad class of in-control distributions would be
extremely useful in statistical practice. Lombard and Van Zyl (2018) developed
distribution-free CUSUMs that can detect deviations from a specified median in
a symmetric continuous distribution. The scale parameter does not figure
directly in the construction of these CUSUMs and no moment conditions are
required to assure their validity. This development frees one from the
restriction to an underlying normal distribution and the difficulties in
designing CUSUMs that operate efficiently when the variance is estimated or
the distribution is non--normal; see, for instance, Bagshaw and Johnson (1974),
Jones et Al. (2004) and Diko et Al. (2020).

Gordon and Pollak (1994) constructed a distribution free S-R type scheme which
they dubbed the NPSR (non parametric Shiryaev-Roberts) scheme, for detecting
shifts away from a known median in a symmetric distribution. The NPSR is based
on the signs of the data and the ranks of their absolute values and\ requires
specification of three adjustable parameters. These permit it being "tuned" to
any given symmetric distribution. However, the NPSR does not seem to have
enjoyed widespread adoption among practitioners. This is possibly due to the
complexity of the scheme which does not allow a simple recursion such as
(\ref{S-R recursion}) and to computational difficulties surrounding the
generation of control limits for a wide range of ICARLs. These difficulties
seem to have thus far inhibited a fuller evaluation of the NPSR's properties.

Following ideas from Lombard and Van Zyl (2018), we propose in this paper an
alternative distribution-free S-R scheme that is based on the signs $s_{i}$ of
the observations and on the \textit{sequential} ranks $r_{i}^{+}$ of their
absolute values. The sequential rank $r_{i}^{+}$ of $|X_{i}|$ in the sequence
$|X_{1}|,\ldots,|X_{i}|$ is the number of observations that are less than or
equal to $|X_{i}|$,%
\[
r_{i}^{+}=1+%
%TCIMACRO{\tsum \nolimits_{j=1}^{i-1}}%
%BeginExpansion
{\textstyle\sum\nolimits_{j=1}^{i-1}}
%EndExpansion
\boldsymbol{1}\left(  |X_{j}|<|X_{i}|\right)  ,\ i\geq2
\]
where $\boldsymbol{1}\left(  \cdot\right)  $ denotes the indicator function
and $r_{1}^{+}=1$. Under any continuous in-control distribution, successive
sequential ranks are mutually independent and uniformly distributed on the
integers $1,2,\ldots i$, hence are distribution free - see Barndorff-Nielsen
(1963, Theorem 1.1). Furthermore, the $r_{i}^{+}$ sequence is then also
independent of the sequence of signs
\[
s_{i}=\boldsymbol{1}\left(  X_{i}>0\right)  -\boldsymbol{1}\left(
X_{i}<0\right)
\]
see Reynolds (1975, Theorem 2.1). Thus, upon replacing $X_{i}$ in
(\ref{S-R recursion}) by a function $J(R_{i}^{s})$ of the signed sequential
ranks%
\begin{equation}
R_{i}^{s}=\frac{s_{i}r_{i}^{+}}{i+1}, \label{seq rank}%
\end{equation}
a new family of distribution-free S-R type change detection schemes appears.
In view of their distribution free property, tables of control limits are
easily generated by Monte Carlo simulation. Furthermore, by an appropriate
choice of the function $J$, the scheme can be tuned to have good properties in
any given distribution. For instance, if the in-control distribution is
expected to be near normal, then upon replacing the $X_{i}$ in
(\ref{S-R recursion}) by
\begin{equation}
X_{i}^{\ast}=\Phi^{-1}((1+R_{i}^{s})/2), \label{VdW}%
\end{equation}
where $\Phi^{-1}$ denotes the inverse of the standard normal cdf, there
results a distribution-free S-R scheme that can be expected to be competitive
with the parametric version when the data indeed come from a normal
distribution. A key fact is that the in-control distribution of $(1+R_{i}%
^{s})/2$ approximates a uniform distribution as $i$ increases, whence the
distribution of $X_{i}^{\ast}$ approximates that of a normally distributed
$X_{i}$. Furthermore, the out-of-control distribution of $X_{i}^{\ast}$
approximates a shifted normal distribution. This construction is reminiscent
of the manner in which the Van der Waerden two sample signed-rank statistic is
obtained - see H\'{a}jek, \v{S}{i}d\'{a}k and Sen (1999, page 118).

The paper is structured as follows. In Section 2 the original Shiryaev-Roberts
scheme and the Gordon and Pollak (1994) NPSR scheme are discussed. In Section
3, we introduce the signed sequential rank S-R schemes, hereafter referred to
as SSR S-R schemes. These schemes are constructed specifically with a view to
detecting a change away from a \textit{known} median in an unspecified
\textit{symmetric} distribution, thus generalizing the normal distribution
based S-R scheme. Tables of control limits guaranteeing a nominal in-control
ARL are provided. In Section 4 we discuss the specification of an appropriate
reference value $\zeta$ in the sequential rank schemes and in Section 5 we
deal in some detail with the out-of-control run length properties of the
Wilcoxon SSR S-R scheme, that is, the SSR S-R scheme based on the Wilcoxon
score $J(u)=u$. In particular we indicate how its out-of-control behaviour may
be assessed on a theoretical basis by using both formal and informal
calculations. It is seen that the Wilcoxon SSR S-R scheme is quite efficient
in normal distributions and can usefully serve as an omnibus distribution-free
scheme. The results of some pertinent Monte Carlo simulations are reported in
Section 6. Specifically, the performance of the Wilcoxon SSR S-R scheme is
compared to that of the Wilcoxon SSR CUSUM and the NPSR scheme. Finally, in
Section 7, implementation of the schemes is illustrated by application to a
new data set that has not been considered in the literature before. In Section
8 we provide a summary of our main results and conclusions and indicate some areas for further research.

\section{The Shiryaev-Roberts and Gordon-Pollak schemes}

An important aspect of any sequential change detection scheme is the extent to
which the realized ICARL agrees with the nominal value, which we denote
generically by $ARL_{0}$. Successful implementation of the normal S-R scheme
(\ref{S-R recursion}) requires that $\sigma$ be known and that the underlying
distribution be normal. We now examine the extent to which deviations from
these assumptions affect its in-control behaviour. Suppose first that $\sigma$
$(=1)$ is unknown to the analyst and that a Phase I estimated standard
deviation $\hat{\sigma}_{m}$, computed from $m$ observations, is used as a
proxy for $\sigma$. Then the analyst will rescale the Phase II data, replacing
$X_{i}$ by $X_{i}/\hat{\sigma}_{m}$, and run the S-R scheme on the rescaled
data. Consequently, since $X_{i}/\hat{\sigma}_{m}$ does not have unit
variance, the ICARL will differ from the nominal $ARL_{0}$. To illustrate,
suppose $ARL_{0}=500$ is desired and that an estimate $\hat{\sigma}_{50}=1.1$
has been found from some Phase I data. Application of the S-R scheme to the
rescaled data will then produce an ICARL of $1081$ (estimated from $10^{5}$
Monte Carlo trials). To provide some context to this result we note that when
$\sigma=1$,
\[
Pr\left[  \hat{\sigma}_{50}>1.1\right]  =Pr\left[  \hat{\sigma}_{50}%
<0.9\right]  =0.125,
\]
which means that about one in every four estimates will be larger that $1.1$
or smaller than $0.9$. Table 1a shows Monte Carlo estimated true ICARL values
for $m=50$, two estimates $\hat{\sigma}_{50}$ of $\sigma$, two reference
values $\zeta$ and three $ARL_{0}$ values.

\begin{center}%
\begin{tabular}
[c]{ll}%
Table 1a & Estimated ICARL when $\sigma$ is\\
& estimated from $50$ observations.
\end{tabular}

$%
\begin{tabular}
[c]{ll|c|c|c|}\cline{3-5}
&  & \multicolumn{3}{|c|}{$ARL_{0}$}\\\hline
\multicolumn{1}{|l}{$\hat{\sigma}_{50}$} & \multicolumn{1}{|c|}{$\zeta$} &
100 & 500 & 1,000\\\hline
\multicolumn{1}{|l}{1.1} & \multicolumn{1}{|c|}{0.25} & \textbf{130} &
\textbf{875} & \textbf{1,983}\\\hline
\multicolumn{1}{|l}{1.1} & \multicolumn{1}{|c|}{0.5} & \textbf{158} &
\textbf{1,081} & \textbf{2,495}\\\hline
\multicolumn{1}{|l}{0.9} & \multicolumn{1}{|c|}{0.25} & \textbf{75} &
\textbf{305} & \textbf{544}\\\hline
\multicolumn{1}{|l}{0.9} & \multicolumn{1}{|c|}{0.5} & \textbf{64} &
\textbf{245} & \textbf{424}\\\hline
\end{tabular}
\ $
\end{center}

The sizes of the discrepancies between nominal and true ICARLs are clearly a
cause for concern. To ameliorate the situation one can increase the size of
the sample on which the estimate is based. However, doubling the number of
Phase I observations does not improve the situation much - see Table 1b, in
which%
\[
Pr\left[  \hat{\sigma}_{100}>1.08\right]  =Pr\left[  \hat{\sigma}%
_{50}<0.92\right]  =0.125.
\]

\begin{center}%
\begin{tabular}
[c]{ll}%
Table 1b & Estimated ICARL when $\sigma$ is\\
& estimated from $100$ observations.
\end{tabular}

$%
\begin{tabular}
[c]{ll|c|c|c|}\cline{3-5}
&  & \multicolumn{3}{|c|}{$ARL_{0}$}\\\hline
\multicolumn{1}{|l}{$\hat{\sigma}_{100}$} & \multicolumn{1}{|c|}{$\zeta$} &
100 & 500 & 1,000\\\hline
\multicolumn{1}{|l}{1.08} & \multicolumn{1}{|c|}{0.25} & \textbf{125} &
\textbf{764} & \textbf{1,723}\\\hline
\multicolumn{1}{|l}{1.08} & \multicolumn{1}{|c|}{0.5} & \textbf{147} &
\textbf{923} & \textbf{2,066}\\\hline
\multicolumn{1}{|l}{0.92} & \multicolumn{1}{|c|}{0.25} & \textbf{80} &
\textbf{334} & \textbf{609}\\\hline
\multicolumn{1}{|l}{0.92} & \multicolumn{1}{|c|}{0.5} & \textbf{70} &
\textbf{282} & \textbf{507}\\\hline
\end{tabular}
\ $
\end{center}

Suppose next that the variance is known to equal $1$ but that the in-control
distribution is non-normal. The last three columns in Table 2 show Monte Carlo
estimates of the ICARL values of the normal distribution S-R scheme when the
data actually come from logistic, Laplace and Student $t_{3}$ distributions,
all standardised to unit variance. Each of the estimates was made on $10^{5}$
Monte Carlo trials. In the table, the first three columns show the reference
values $\zeta=0.1,$ $0.15,$ $0.5$ and $0.75$, the nominal ICARL values
$ARL_{0}=500$ and $ARL_{0}=1,000$ and the corresponding control limits $h$
that guarantee an ICARL equal to $ARL_{0}$ under a normal distribution.

The differences between the estimated true in-control ARLs and the nominal
$ARL_{0}$ values could be considered acceptable "for practical purposes" only
in the logistic and Laplace cases at $\zeta=0.1$, which is a reference value
that would not be used frequently. The results shown in Tables 1a and 1b and
in Table 2 clearly indicate that distribution-free and scale invariant S-R
schemes would be valuable additions to the toolbox of a practitioner who
contemplates using an S-R type scheme.

\begin{center}%
\begin{tabular}
[c]{ll}%
Table 2 & Comparison of estimated true and nominal\\
& ICARL of Shiryaev-Roberts scheme\\
& in non-normal distributions.
\end{tabular}

${\footnotesize
\begin{tabular}
[c]{ccc|c|c|c|}\cline{4-6}
&  &  & \multicolumn{3}{|c|}{Estimated ICARL}\\\hline
\multicolumn{1}{|c}{$\zeta$} & \multicolumn{1}{|c}{$ARL_{0}$} &
\multicolumn{1}{|c|}{$h$} & Logistic & Laplace & $t_{3}$\\\hline
\multicolumn{1}{|c}{0.1} & \multicolumn{1}{|c}{500} &
\multicolumn{1}{|c|}{6.10} & \textbf{512} & \textbf{513} & \textbf{581}%
\\\hline
\multicolumn{1}{|c}{} & \multicolumn{1}{|c}{1,000} &
\multicolumn{1}{|c|}{6.79} & \textbf{1007} & \textbf{1020} & \textbf{1156}%
\\\hline
\multicolumn{1}{|c}{0.25} & \multicolumn{1}{|c}{500} &
\multicolumn{1}{|c|}{5.92} & \textbf{484} & \textbf{483} & \textbf{560}%
\\\hline
\multicolumn{1}{|c}{} & \multicolumn{1}{|c}{1,000} &
\multicolumn{1}{|c|}{6.62} & \textbf{961} & \textbf{946} & \textbf{970}%
\\\hline
\multicolumn{1}{|c}{0.5} & \multicolumn{1}{|c}{500} &
\multicolumn{1}{|c|}{5.63} & \textbf{418} & \textbf{353} & \textbf{362}%
\\\hline
\multicolumn{1}{|c}{} & \multicolumn{1}{|c}{1,000} &
\multicolumn{1}{|c|}{6.32} & \textbf{804} & \textbf{622} & \textbf{527}%
\\\hline
\multicolumn{1}{|c}{0.75} & \multicolumn{1}{|c}{500} &
\multicolumn{1}{|c|}{5.35} & \textbf{330} & \textbf{244} & \textbf{242}%
\\\hline
\multicolumn{1}{|c}{} & \multicolumn{1}{|c}{1,000} &
\multicolumn{1}{|c|}{6.06} & \textbf{598} & \textbf{392} & \textbf{334}%
\\\hline
\end{tabular}
}$
\end{center}

To the best of our knowledge the first, and to date only, distribution-free
S-R type scheme applicable to symmetric distributions with known median is the
NPSR of Gordon and Pollak (1994). The NPSR is based on a double sequence
$\Lambda_{k}^{n},\ 1\leq k\leq n,\ n\geq1$, of nonlinear two-sample rank and
sign statistics, not on signed ranks alone. An expression for $\Lambda_{k}%
^{n}$, which has a somewhat complicated structure, can be found in eqn. (6) of
Gordon and Pollak (1994) who also provide efficient Matlab code for the
calculation. The run length is the first index $n$ at which the sum $R_{n}=%
%TCIMACRO{\tsum \nolimits_{k=1}^{n}}%
%BeginExpansion
{\textstyle\sum\nolimits_{k=1}^{n}}
%EndExpansion
\Lambda_{k}^{n}$ exceeds a control limit, $A$, which makes the ICARL equal to
a specified number $ARL_{0}$. Given a large nominal $ARL_{0}$, Gordon and
Pollak (1994) provide an approximation to $A$ in terms of $ARL_{0}$ and a
further three adjustable parameters. However, the approximation needs to be
supplemented by Monte Carlo simulations to\ find a value of $A$ that produces
an ICARL sufficiently close to $ARL_{0}$ for use in a practical application.
Here one encounters two kinds of difficulties. The first difficulty is that
the complicated structure of $R_{n}$ results in excessively time consuming
simulations at large $ARL_{0}$ values. The second, and perhaps more important,
difficulty is that a small fraction of simulated run lengths are so large that
the only way in which a practicable scheme results is if the run length is
truncated. Gordon and Pollak truncated all run lengths at $2,500$. In our
simulations we truncated all NPSR run lengths at $5,000$ and encountered a
negligible proportion of NPSR run lengths that had to be truncated.

\section{Signed sequential rank schemes}

The independence, distribution freeness and naturally sequential nature of
signed sequential ranks $R_{i}^{s}$ from (\ref{seq rank}) makes them ideally
suited to the construction of CUSUM and S-R schemes for independently
distributed time ordered data. A class of signed sequential rank analogues of
(\ref{Cusum recursion}) and (\ref{S-R recursion}) is obtained upon replacing
$X_{i}$ there by
\begin{equation}
X_{i}^{\ast}=\frac{J\left(  R_{i}^{s}\right)  }{v_{i}} \label{SSRxi}%
\end{equation}
where $J(u),\ -1<u<1$ is an odd, square-integrable, function on the interval
$(-1,1)$ and where
\begin{equation}
v_{i}=\sqrt{\frac{1}{i}\sum\nolimits_{j=1}^{i}J^{2}\left(  \frac{j}%
{i+1}\right)  }. \label{score var}%
\end{equation}

It is customary in the rank statistic literature to refer to $J(u)$ as a score
function. In particular, the Wilcoxon score
\[
J_{W}(u)=u,\ -1\leq u\leq1
\]
leads to a particularly useful omnibus scheme. In this case%
\[
\nu_{i}=\sqrt{6(i+1)/(2i+1)}%
\]
and the corresponding SSR S-R and CUSUM schemes are defined by%
\begin{equation}
{D}_{i}=\log(1+\exp\left(  {D}_{i-1}\right)  )+2\zeta\left(  \frac{R_{i}^{s}%
}{\nu_{i}}-\zeta\right)  \label{Wilcoxon S-R}%
\end{equation}
and
\begin{equation}
C_{i}=\max\left(  0,C_{i-1}+\frac{R_{i}^{s}}{\nu_{i}}-\zeta\right)  .
\label{Wilcoxon CUSUM}%
\end{equation}
For underlying distributions close to the normal, one could use an SSR scheme
based on the Van der Waerden score already mentioned in the introduction,
namely
\[
J_{V}(u)=\Phi^{-1}((1+u)/2).
\]
However, the Pearson correlation coefficient between $J_{W}(R_{i}^{s})$ and
$J_{V}(R_{i}^{s})$ tends as $i\rightarrow\infty$ to $\sqrt{\pi/3}=0.98$, which
implies that not much will be lost if the computationally convenient $J_{W}$
is used in place of the somewhat more complicated $J_{V}$. Furthermore the
large sample correlations between the Wilcoxon score and the efficient scores
in some heavy-tailed Student $t$-distributions are quite high while their
correlations with the Van der Waerden score are somewhat lower. Table 3 shows
these correlations in four $t$-distributions.

\begin{center}%
\begin{tabular}
[c]{ll}%
Table 3 & Correlations of efficient scores in\\
& Student $t_{\nu}$ distributions with Wilcoxon\\
& and Van der Waerden scores.
\end{tabular}

{\footnotesize
\begin{tabular}
[c]{c|c|c|c|c|c|}\cline{2-6}
& \multicolumn{5}{|c|}{Distribution}\\\hline
\multicolumn{1}{|c|}{Score} & $normal$ & $t_{4}$ & $t_{3}$ & $t_{2}$ & $t_{1}
$\\\hline
\multicolumn{1}{|c|}{Wilcoxon} & 0.98 & 0.99 & 0.98 & 0.94 & 0.79\\\hline
\multicolumn{1}{|c|}{Van der Waerden} & 1.00 & 0.95 & 0.92 & 0.86 &
0.67\\\hline
\end{tabular}
}
\end{center}

Consequently, the SSR S-R and SSR CUSUM schemes that are based on the
Wilcoxon score $J_{W}$ can be expected to be quite efficient in a broad range
of symmetric underlying distributions, obviating to a large extent the
necessity of tuning the scheme to any specific distribution. We will refer to
them as the Wilcoxon SSR S-R and the Wilcoxon SSR CUSUM respectively. The
latter one of these was dealt with in detail by Lombard and Van Zyl (2018).
The term VdW SSR S-R will be used to denote the SSR S-R scheme that is based
on the Van der Waerden score $J_{V}$. Construction of these schemes does
not require knowledge of the numerical value of any scale parameter $\sigma$
because signed sequential ranks are scale invariant.

Tables 4 and 5 give control limits for a matrix of $(\zeta,ARL_{0})$ pairs for
the Wilcoxon SSR S-R and CUSUM schemes. The tables were generated by Monte
Carlo simulation using the method detailed in Lombard and Van Zyl
(2018).\newpage

\begin{center}%
\begin{tabular}
[c]{ll}%
Table 4 & Control limits for the Wilcoxon SSR S-R .
\end{tabular}

{\footnotesize
\begin{tabular}
[c]{c|c|c|c|c|c|c|c|}\cline{2-8}
& \multicolumn{7}{|c|}{$ARL_{0}$}\\\hline
\multicolumn{1}{|c|}{$\zeta$} & 100 & 200 & 300 & 400 & 500 & 1,000 &
2,000\\\hline
\multicolumn{1}{|c|}{0.05} & 4.55 & 5.24 & 5.65 & 5.93 & 6.16 & 6.85 &
7.55\\\hline
\multicolumn{1}{|c|}{0.10} & 4.49 & 5.18 & 5.60 & 5.86 & 6.10 & 6.80 &
7.48\\\hline
\multicolumn{1}{|c|}{0.15} & 4.43 & 5.14 & 5.53 & 5.83 & 6.05 & 6.73 &
7.24\\\hline
\multicolumn{1}{|c|}{0.20} & 4.37 & 5.07 & 5.47 & 5.76 & 5.98 & 6.68 &
7.38\\\hline
\multicolumn{1}{|c|}{0.25} & 4.31 & 5.01 & 5.41 & 5.70 & 5.92 & 6.59 &
7.27\\\hline
\multicolumn{1}{|c|}{0.30} & 4.29 & 4.95 & 5.34 & 5.61 & 5.83 & 6.51 &
7.18\\\hline
\multicolumn{1}{|c|}{0.35} & 4.21 & 4.86 & 5.28 & 5.55 & 5.74 & 6.41 &
7.08\\\hline
\multicolumn{1}{|c|}{0.40} & 4.13 & 4.78 & 5.17 & 5.43 & 5.66 & 6.31 &
6.96\\\hline
\multicolumn{1}{|c|}{0.45} & 4.04 & 4.68 & 5.07 & 5.33 & 5.54 & 6.18 &
6.83\\\hline
\multicolumn{1}{|c|}{0.50} & 3.95 & 4.58 & 4.95 & 5.24 & 5.43 & 6.03 &
6.69\\\hline
\multicolumn{1}{|c|}{0.75} & 3.38 & 3.95 & 4.26 & 4.50 & 4.67 & 5.22 &
5.77\\\hline
\multicolumn{1}{|c|}{1.00} & 2.65 & 3.09 & 3.36 & 3.55 & 3.69 & 4.12 &
4.54\\\hline
\end{tabular}
\bigskip}%

\begin{tabular}
[c]{ll}%
Table 5 & Control limits for the Wilcoxon SSR CUSUM.
\end{tabular}

{\footnotesize
\begin{tabular}
[c]{c|c|c|c|c|c|c|c|}\cline{2-8}
& \multicolumn{7}{|c|}{${\small ARL}_{0}$}\\\hline
\multicolumn{1}{|c|}{${\small \zeta}$} & 100 & 200 & 300 & 400 & 500 & 1,000 &
2,000\\\hline
\multicolumn{1}{|c|}{0.00} & 8.92 & 13.07 & 16.24 & 18.9 & 21.3 & 30.24 &
43.95\\\hline
\multicolumn{1}{|c|}{0.05} & 7.61 & 10.51 & 12.49 & 14.01 & 15.33 & 19.89 &
25.03\\\hline
\multicolumn{1}{|c|}{0.10} & 6.45 & 8.62 & 10.05 & 11.12 & 12.01 & 14.79 &
17.93\\\hline
\multicolumn{1}{|c|}{0.15} & 5.65 & 7.34 & 8.42 & 9.21 & 9.86 & 11.88 &
14.06\\\hline
\multicolumn{1}{|c|}{0.20} & 5.00 & 6.37 & 7.24 & 7.87 & 8.37 & 9.96 &
11.57\\\hline
\multicolumn{1}{|c|}{0.25} & 4.46 & 5.61 & 6.33 & 6.85 & 7.25 & 8.52 &
9.84\\\hline
\multicolumn{1}{|c|}{0.30} & 4.01 & 5.00 & 5.60 & 6.03 & 6.37 & 7.45 &
8.53\\\hline
\multicolumn{1}{|c|}{0.35} & 3.62 & 4.48 & 5.00 & 5.37 & 5.66 & 6.58 &
7.51\\\hline
\multicolumn{1}{|c|}{0.40} & 3.29 & 4.04 & 4.49 & 4.81 & 5.06 & 5.87 &
6.66\\\hline
\multicolumn{1}{|c|}{0.45} & 2.99 & 3.66 & 4.05 & 4.34 & 4.56 & 5.25 &
5.96\\\hline
\multicolumn{1}{|c|}{0.50} & 2.73 & 3.31 & 3.68 & 3.93 & 4.13 & 4.74 &
5.34\\\hline
\multicolumn{1}{|c|}{0.75} & 1.72 & 2.06 & 2.2 & 2.42 & 2.53 & 2.89 &
3.25\\\hline
\multicolumn{1}{|c|}{1.00} & 1.02 & 1.24 & 1.34 & 1.42 & 1.49 & 1.71 &
1.92\\\hline
\end{tabular}
}\newpage
\end{center}

\section{Specification of the reference value}

Consider a situation in which the median shifts after a considerable time
$\tau$ from $0$ to $\delta\neq0$. Then, with
\[
\xi_{i}=\frac{R_{i}^{s}}{\nu_{i}}=\frac{s_{i}r_{i}^{+}}{v_{i}(i+1)},
\]
a calculation which is detailed in the Appendix shows that for $\tau$ large
and $\delta$ small,
\begin{equation}
E[\xi_{\tau+1}]\approx\theta_{0}\delta, \label{OOCmean_ksi}%
\end{equation}
with
\begin{equation}
\theta_{0}=\sqrt{12}\int_{-\infty}^{\infty}f^{2}(x)dx, \label{thetazeroSSR}%
\end{equation}
$f$ denoting the pdf of the in-control distribution. In analogy with the
parametric scheme (\ref{S-R recursion}), the relation (\ref{OOCmean_ksi})
suggests $\zeta=\delta_{1}\theta_{0}/2$ as an appropriate choice for targeting
a shift of size $\delta_{1}$. Some values of $\theta_{0}$ that are likely to
be encountered in practice are given in Table 6. In the case of the $t_{2}$
and $t_{1}\ $(Cauchy) distributions the interquartile range served as the
scale parameter, that is, the shift $\delta$ is expressed in units of the
interquartile range. Even though the density function underlying the data is
unknown, we can still use these values of $\theta_{0}$ to make a priori an
informed choice of $\theta_{0}$, hence of $\zeta$, based on the expected
heaviness of the tails. On the other hand, if some Phase I data are available,
$\theta_{0}$ can be estimated non-parametrically as indicated in Lombard and
Van Zyl (2018, Section 3.1).

\begin{center}%
\begin{tabular}
[c]{ll}%
Table 6 & Values of $\theta_{0}$ for a range of symmetric distributions.
\end{tabular}

${\footnotesize
\begin{tabular}
[c]{|c|c|c|c|c|c|}\hline
\multicolumn{6}{|c|}{Distribution}\\\hline
normal & Laplace & $t_{4}$ & $t_{3}$ & $t_{2}$ & $t_{1}$\\\hline
$0.98$ & $1.2$ & $1.18$ & $1.37$ & $1.18$ & $1.10$\\\hline
\end{tabular}
}$
\end{center}

\section{Run length properties}

An important practical requirement is to assess a priori the properties of the
run length%
\begin{equation}
N=min\left\{  i\geq1;\ D_{i}\geq h\right\}  \label{S-R run length}%
\end{equation}
with $D_{i}$ from (\ref{Wilcoxon S-R}). Given any specific out-of-control
density or range of densities, the behaviour of $N$ can be assessed quite
simply by Monte Carlo simulation. The question nevertheless arises whether it
is possible to obtain useful conclusions based on less information. To see
that this is indeed possible, we observe that $D_{i}$ can be expressed as a
functional of the partial sums $S_{i}=\xi_{1}+\cdots+\xi_{i}$, namely%
\[
exp\left(  D_{i}\right)  =%
%TCIMACRO{\tsum \nolimits_{j=1}^{i-1}}%
%BeginExpansion
{\textstyle\sum\nolimits_{j=1}^{i-1}}
%EndExpansion
exp\left\{  2\zeta\left(  S_{i}-S_{j}\right)  -\zeta^{2}(i-j)\right\}
\]
- see, for instance, Pollak and Siegmund (1991, page 396). Because the partial
sums $S_{i}$ are for large $i$ approximately normally distributed, we can
expect the behaviour of a Wilcoxon SSR S-R at a small out-of-control shift
$\delta$ to be close to that of a normal S-R scheme (\ref{S-R recursion}) with
the same $\zeta$ and $h$ at a shift $\theta_{0}\delta$, provided $ARL_{0}$ and
the change point $\tau$ are large - see (\ref{OOCmean_ksi}) and
(\ref{thetazeroSSR}). Large values of $ARL_{0}$, hence $h$, and $\tau$ allows
enough time for a normal approximation to manifest itself. Indeed, Lombard and
Van Zyl (2018, Appendix) showed via a continuous time approximation involving
Brownian motion that under these conditions the ARL of any SSR CUSUM behaves
in the indicated manner. They also provided supporting numerical evidence -
see their Tables 4.1 and 4.2 and Tables S3 and S4 in the Supplementary
Material. The same method substantiates the conclusion in respect of general
SSR S-R schemes and the following numerical evidence provides support
specifically in respect of the Wilcoxon SSR S-R scheme.

The out-of-control performance criterion we use here is the conditional
average delay time (CADT) $E_{\tau}[N-\tau|N>\tau]$, where the subscript
$\tau$ in $E_{\tau}$ denotes that the expected value is computed under the
assumption that the change occurs at time $t=\tau+1$. The $ARL_{0}$ value was
set at $500$, the target out of control shifts were $\delta_{1}=0.2$ and
$\delta_{1}=0.5$. The Laplace distribution and the much heavier tailed $t_{3}$
distributions both served as in-control distributions. The design parameters
$\theta_{0}$ (the tuning constant) and $(\zeta,h)=(\delta_{1}\theta_{0}%
/2,h)=$(reference constant,control limit) are shown in the second row of
Tables 7a and 7b. The entries in the tables are Monte Carlo estimates
($10^{5}$ trials) of the conditional average delays, $\mathcal{W}(\delta
)\ $and $\mathcal{N}(\delta\theta_{0}\mathcal{)}$, of the Wilcoxon SSR S-R and
normal S-R schemes respectively at a range of out-of-control means $\delta$.
The results are shown for changes of size $\delta$ at change points $\tau
=0\ $and $\tau=100$. If the out-of-control behaviour of an SSR S-R scheme is
indeed similar to that of a normal S-R scheme, using normal data only, with a $\theta_{0}$-adjusted
out-of-control shift, we would expect to see
\begin{equation}
\mathcal{W}(\delta)\approx\mathcal{N}(\delta\theta_{0})
\label{OOCnormalApprox}%
\end{equation}
to good approximation if $\delta$ is "small" and $\tau$ is "large". Inspection
of the results in Table 7a indicates that the approximation is quite
satisfactory at $\tau=100$. Also, though rather unexpectedly, the
approximation is also quite good at $\tau=0$ (Table 7b) except at $\delta
\geq0.5$, that is, when the underlying process is substantially out-of-control
from the outset.

\begin{center}%
\begin{tabular}
[c]{ll}%
{\small Table 7a} & {\small Wilcoxon SSR S-R CADT approximations in Laplace
and }\\
& $t_{3}$ {\small distributions. }$ARL_{0}=500${\small ; change point }%
$\tau=100${\small.}%
\end{tabular}

$%
\begin{tabular}
[c]{c|c|c|c|c|c|c|c|c|}\cline{2-9}
& \multicolumn{4}{|c|}{{\small Laplace: (}$\theta_{0}=1.2)$} &
\multicolumn{4}{|c|}{$t_{3}${\small : (}$\theta_{0}=1.37)$}\\\hline
\multicolumn{1}{|c|}{${\small (\zeta,h)}$} &
\multicolumn{2}{|c|}{${\small (0.12,6.09)}$} &
\multicolumn{2}{|c|}{${\small (0.3,5.83)}$} &
\multicolumn{2}{|c|}{${\small (0.15,6.05)}$} &
\multicolumn{2}{|c|}{${\small (0.35,5.74)}$}\\\hline
\multicolumn{1}{|c|}{${\small \delta}$} & ${\small W(\delta)}$ &
${\small N(\theta}_{0}{\small \delta)}$ & ${\small W(\delta)}$ &
${\small N(\theta}_{0}{\small \delta)}$ & ${\small W(\delta)}$ &
${\small N(\theta}_{0}{\small \delta)}$ & ${\small W(\delta)}$ &
${\small N(\theta}_{0}{\small \delta)}$\\\hline
\multicolumn{1}{|c|}{{\small 0.125}} & {\small 91} & {\small 95} &
{\small 124} & {\small 123} & {\small 85} & {\small 85} & {\small 116} &
{\small 118}\\\hline
\multicolumn{1}{|c|}{{\small 0.25}} & {\small 44} & {\small 45} & {\small 49}
& {\small 50} & {\small 38} & {\small 37} & {\small 43} & {\small 42}\\\hline
\multicolumn{1}{|c|}{{\small 0.5}} & {\small 22} & {\small 21} & {\small 19} &
{\small 18} & {\small 18} & {\small 17} & {\small 16} & {\small 15}\\\hline
\multicolumn{1}{|c|}{{\small 0.75}} & {\small 15} & {\small 14} & {\small 12}
& {\small 11} & {\small 12} & {\small 11} & {\small 10} & {\small 8}\\\hline
\multicolumn{1}{|c|}{{\small 1.0}} & {\small 12} & {\small 11} & {\small 10} &
{\small 8} & {\small 10} & {\small 8} & {\small 8} & {\small 6}\\\hline
\multicolumn{1}{|c|}{{\small 1.5}} & {\small 10} & {\small 7} & {\small 7} &
{\small 7} & {\small 8} & {\small 6} & {\small 6} & {\small 5}\\\hline
\end{tabular}
\ $%

\begin{tabular}
[c]{ll}%
{\small Table 7b} & {\small Wilcoxon SSR S-R CADT approximations in Laplace
and}\\
& $t_{3}$ {\small distributions. }$ARL_{0}=500${\small ; change point }$\tau
=0${\small .}%
\end{tabular}

$%
\begin{tabular}
[c]{c|c|c|c|c|c|c|c|c|}\cline{2-9}
& \multicolumn{4}{|c|}{{\small Laplace: (}$\theta_{0}=1.2)$} &
\multicolumn{4}{|c|}{$t_{3}${\small : (}$\theta_{0}=1.37)$}\\\hline
\multicolumn{1}{|c|}{${\small (\zeta,h)}$} &
\multicolumn{2}{|c|}{${\small (0.12,6.09)}$} &
\multicolumn{2}{|c|}{${\small (0.3,5.83)}$} &
\multicolumn{2}{|c|}{${\small (0.15,6.05)}$} &
\multicolumn{2}{|c|}{${\small (0.35,5.74)}$}\\\hline
\multicolumn{1}{|c|}{${\small \delta}$} & ${\small W(\delta)}$ &
${\small N(\theta}_{0}{\small \delta)}$ & ${\small W(\delta)}$ &
${\small N(\theta}_{0}{\small \delta)}$ & ${\small W(\delta)}$ &
${\small N(\theta}_{0}{\small \delta)}$ & ${\small W(\delta)}$ &
${\small N(\theta}_{0}{\small \delta)}$\\\hline
\multicolumn{1}{|c|}{{\small 0.125}} & {\small 124} & {\small 125} &
{\small 135} & {\small 132} & {\small 107} & {\small 107} & {\small 126} &
{\small 118}\\\hline
\multicolumn{1}{|c|}{{\small 0.25}} & {\small 67} & {\small 65} & {\small 60}
& {\small 56} & {\small 55} & {\small 53} & \textbf{54} & \textbf{47}\\\hline
\multicolumn{1}{|c|}{{\small 0.5}} & \textbf{40} & \textbf{34} & \textbf{27} &
\textbf{22} & \textbf{32} & \textbf{27} & \textbf{24} & \textbf{18}\\\hline
\multicolumn{1}{|c|}{{\small 0.75}} & \textbf{31} & \textbf{24} & \textbf{19}
& \textbf{14} & \textbf{26} & \textbf{18} & \textbf{17} & \textbf{11}\\\hline
\multicolumn{1}{|c|}{{\small 1.0}} & \textbf{28} & \textbf{19} & \textbf{16} &
\textbf{10} & \textbf{23} & \textbf{14} & \textbf{14} & \textbf{8}\\\hline
\multicolumn{1}{|c|}{{\small 1.5}} & \textbf{25} & \textbf{13} & \textbf{14} &
\textbf{7} & \textbf{20} & \textbf{10} & \textbf{12} & \textbf{6}\\\hline
\end{tabular}
$
\end{center}

The preceding results indicate that in the absence of a specified
out-of-control density, useful estimates of out-of-control ARLs can be had if
an estimate of $\theta_{0}$ is available. The behaviour of the ARLs described
above and seen in Tables 7a and 7b can be deduced from formal limit theorems
involving contiguous and fixed alternatives. The approximation
(\ref{OOCnormalApprox}) is an informal interpretation of Theorem 1 in Lombard
(1981) which deals with the convergence in distribution of signed sequential
rank statistics under contiguous alternatives to a drifted Brownian motion.
Here, "contiguous" is interpreted informally as indicating that $\delta$ and
the reference value $\zeta$ are "small"\ when $h$ is "large" - see also the
Appendix in Lombard and Van Zyl (2018). The failure of (\ref{OOCnormalApprox})
when $\tau$ $(=0)$ is "small" and $\delta\ (\geq0.5)$ is "large" is to a large
extent explained by Theorem 1.1 in M\"{u}ller-Funk (1983) and Theorem 2 in
Lombard and Mason (1985) which deal with the convergence in distribution of
sequential rank statistics under \textit{fixed} alternatives. In this case
there is distributional convergence to Gaussian processes, which are not
simply drifted Brownian motions, and for which run length distributions are
not available. As far as we are aware, similar approximation results are not
available for the NPSR.

To summarize, there are strong indications (except when the underlying process
is substantially out of control from the outset) that the out of control
behaviour of the two Wilcoxon SSR schemes can usefully be gauged from the
behaviour of their normal distribution counterparts by the simple device of
replacing in the latter any shift $\delta$ by $\hat{\theta}_{0}\delta$ where
$\hat{\theta}_{0}$ is an estimator of $\theta_{0}$ made from some Phase I
data. Thus, for instance, if an estimate of the CADT at a given shift $\delta$
and change point $\tau$ is required, then this can be had by simulating
observations $X_{1},\ldots,X_{\tau}$ from a normal($0,1$) distribution and
observations $X_{\tau+1},X_{\tau+2},\ldots$ from a normal($\hat{\theta}%
_{0}\delta,1$)$\ $distribution.

\section{Simulation results}

In a numerical comparison of the normal distribution S-R and CUSUM schemes,
Moustakides, Polunchenko and Tartakovsky (2009) conclude that the only marked
difference in out-of-control performances is seen at small shifts. The
approximation argument formulated in Section 5 suggests that this may also be
the major difference between the signed sequential rank analogues of the two
schemes. To investigate this suggestion, we compared in a simulation study the
performances of the Wilcoxon SSR S-R with the Wilcoxon SSR CUSUM of Lombard and Van
Zyl (2018) and with the NPSR of Gordon and Pollak (1994). We report below some
of the results that are indicative of the behaviours of the schemes. When
there is no danger of confusion, we will often refer in what follows to the
S-R, the CUSUM and the NPSR, dropping the "Wilcoxon" and "SSR" prefixes as
well as the "scheme" suffix.

All three schemes were tuned to the detection of mean shifts in two
distributions, namely, the normal distribution and the Laplace distribution.
The use of the Laplace distribution is justified in view of the inclusion of
the NPSR in the comparisons. The NPSR is derived from a mixture of two
exponential distributions, the null instance of which is the Laplace
distribution. To begin with, we limit our reporting to target sizes
$\delta_{1}=0.5$ and $\delta_{1}=0.25$ and $ARL_{0}=500$. In the simulations,
each estimated CADT of the S-R and CUSUM is the result of $10,000$ independent
Monte Carlo trials while the estimates for the NPSR resulted from $5,000$
independent trials. The smaller number of trials used in the case of the NPSR
was necessary in order to keep its runtimes within reasonable bounds. The
somewhat "wobbly" appearance of some of the NPSR plots is a consequence of the
reduced number of trials, but we do not believe that this misrepresents the
true behaviour of the NPSR. In all the figures that follow, the dotted line
represents the S-R, the dashed line represents the NPSR and the solid line the CUSUM.

Beginning with the normal distribution, the tuning parameters used for a
target shift $\delta_{1}=0.5$ were $\zeta=0.25$ for the Wilcoxon SSR S-R and
CUSUM and $(\alpha,\beta,p)=(0.735,1.324,0.691)$ for the NPSR, the latter
found by numerical calculation from equations (10) and (15) in Gordon and
Pollak (1994). The control limits were $h=5.92$ for the S-R, $h=7.25$ for the
CUSUM and $A=375$ for the NPSR. For a target shift $\delta_{1}=0.25$ the
tuning parameters used were $\zeta=0.125$ and $(\alpha,\beta
,p)=(0.860,1.155,0.0.599)\ $with control limits $h=6.07$ for the S-R,
$h=10.94$ for the CUSUM and $A=455$ for the NPSR. Suppose first that the data
do come from a normal distribution. Figures 1 and 2 show plots of the
CADTs against a series of change points
$\tau=0:50:500$ involving actual shifts of sizes $\delta=0.125$ and
$\delta=0.5$. Clearly, the differences between the CADTs of the schemes are
largest when the underlying process is out of control from the outset. Also,
while the three schemes perform similarly when the actual shift is equal to
the target ($\delta=\delta_{1}$, Figure 2), the CUSUM seems to fare rather
poorly compared to the other two when the actual shift is substantially less
than the target ($\delta=\delta_{1}/4$, Figure 1). A striking feature in both
plots is that the CADTs seem to have reached stationary values at or near
change point $\tau\approx50$. This feature also appeared in numerous other
configurations (not shown here).

In Figures 3 and 4 the CADTs at two target shifts $\delta_{1}=0.25$ and
$\delta_{1}=0.5$ are plotted against a series of actual shift sizes occurring
at a change point $\tau=100$. We notice that, as was seen in Figures 1 and 2,
the CADT of all three schemes increases as the actual shift decreases further
away from the target. However, this is not necessarily an indicator of poor
performance because a change of size much less than the target is often deemed
to be undesirable or even as constituting a false alarm. Clearly, the S-R and
NPSR again perform quite similarly at small shifts but with somewhat smaller
CADTs there than the CUSUM.

\begin{center}
$%
\begin{array}
[c]{cc}%
%TCIMACRO{\FRAME{itbpFU}{2.8911in}{2.1975in}{0in}{\Qcb{\QTR{tiny}{Figure
%1}}}{\Qlb{FIG1}}{Figure}{\special{ language "Scientific Word";
%type "GRAPHIC";  display "USEDEF";  valid_file "F";  width 2.8911in;
%height 2.1975in;  depth 0in;  original-width 5.9049in;
%original-height 4.3561in;  cropleft "0";  croptop "1";  cropright "1";
%cropbottom "0";  filename 'FIG_0.eps';file-properties "XNPEU";}} }%
%BeginExpansion
{\parbox[b]{2.8911in}{\begin{center}
\includegraphics[
height=2.1975in,
width=2.8911in
]%
{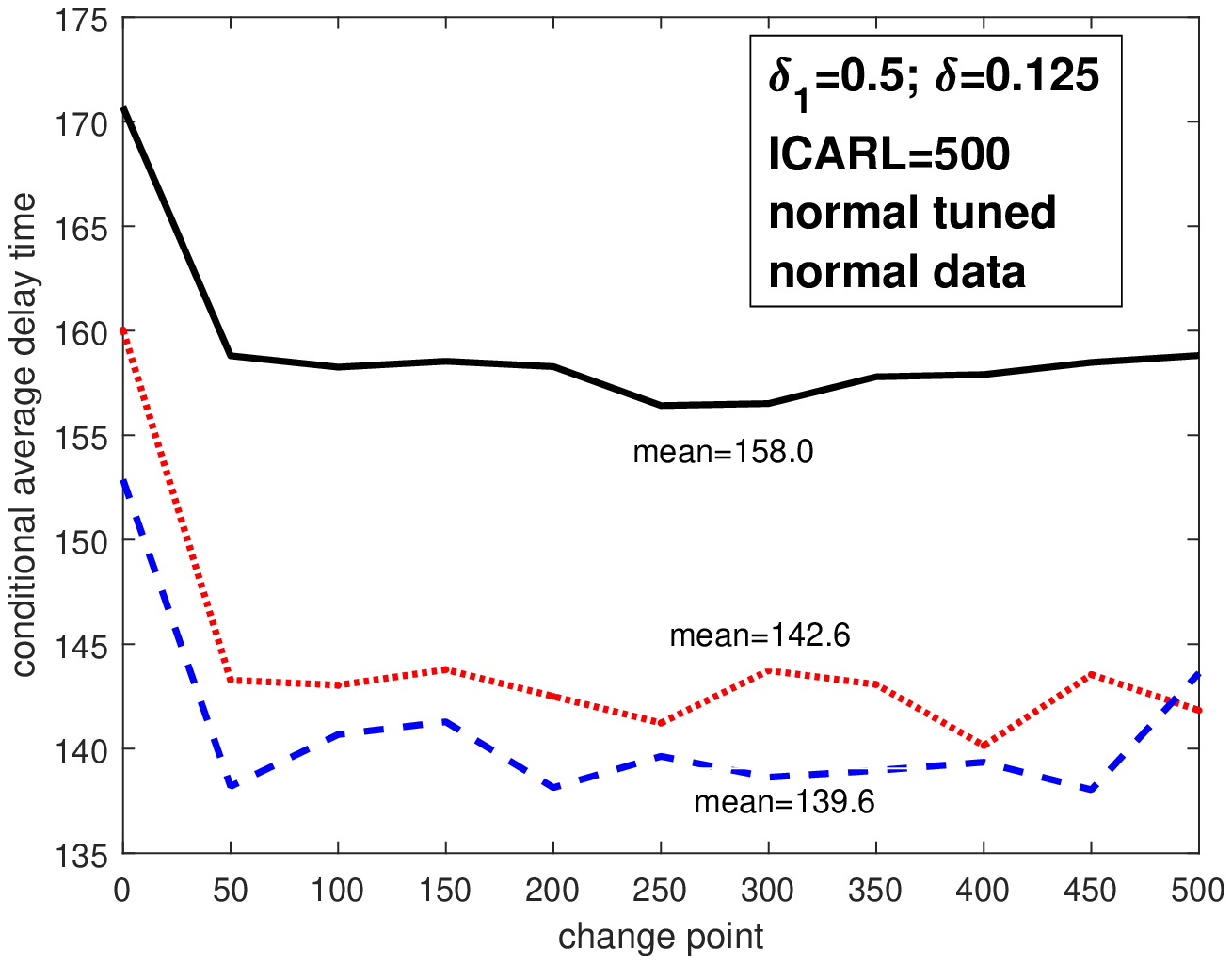}%
\\
{\protect\tiny Figure 1}%
\end{center}}}
%EndExpansion
&
%TCIMACRO{\FRAME{itbpFU}{2.6429in}{2.1205in}{0in}{\Qcb{\QTR{tiny}{Figure
%2}}}{\Qlb{FIG 2}}{Figure}{\special{ language "Scientific Word";
%type "GRAPHIC";  display "USEDEF";  valid_file "F";  width 2.6429in;
%height 2.1205in;  depth 0in;  original-width 5.834in;
%original-height 4.3708in;  cropleft "0";  croptop "1";  cropright "1";
%cropbottom "0";  filename 'FIGURE_2.eps';file-properties "XNPEU";}} }%
%BeginExpansion
{\parbox[b]{2.6429in}{\begin{center}
\includegraphics[
height=2.1205in,
width=2.6429in
]%
{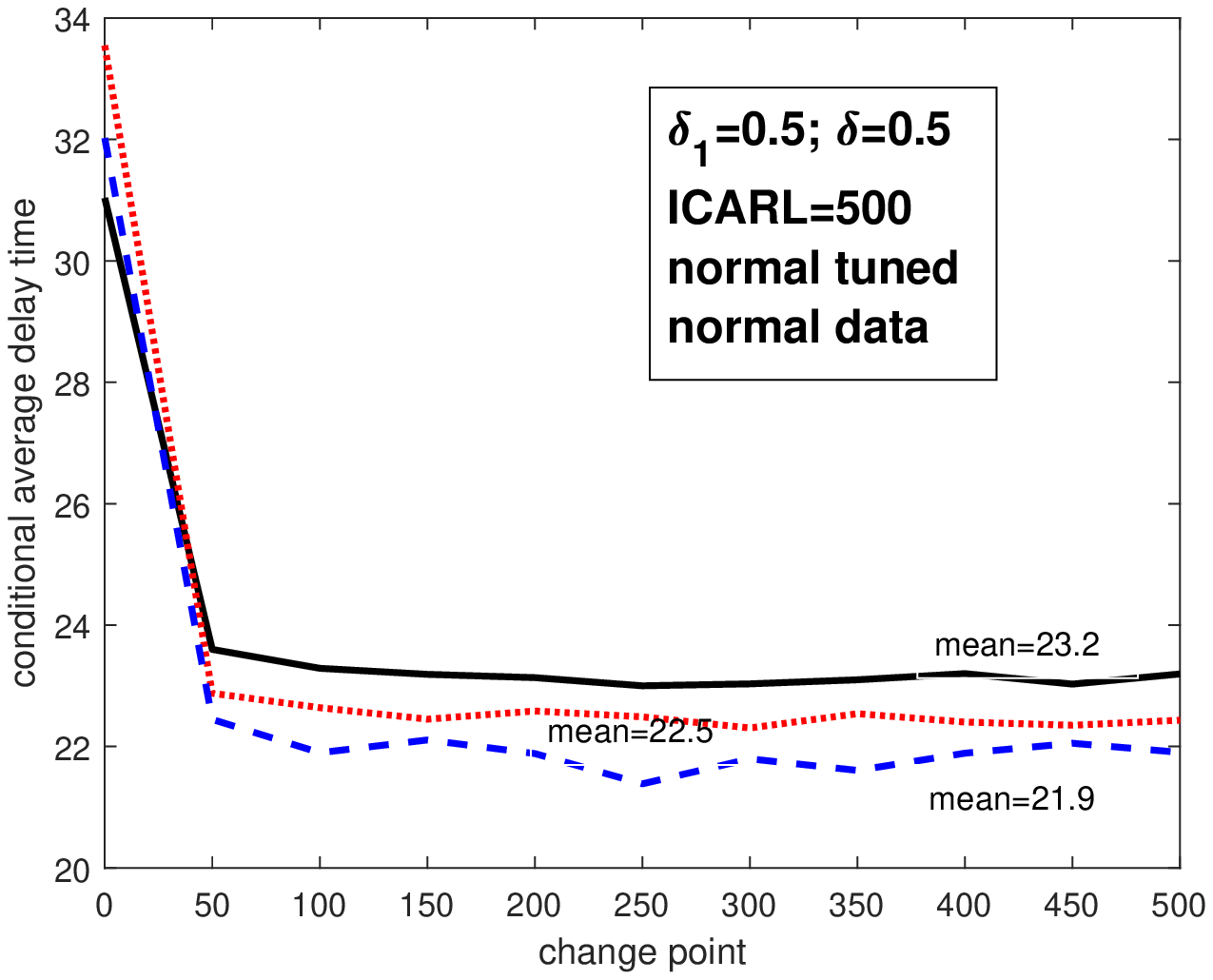}%
\\
{\protect\tiny Figure 2}%
\end{center}}}
%EndExpansion
\\%
%TCIMACRO{\FRAME{itbpFU}{2.5261in}{1.9787in}{0in}{\Qcb{\QTR{tiny}{Figure 3}}}%
%{}{arl_{m}u_{2}.eps}{\special{ language "Scientific Word";  type "GRAPHIC";
%display "USEDEF";  valid_file "F";  width 2.5261in;  height 1.9787in;
%depth 0in;  original-width 5.834in;  original-height 4.3708in;  cropleft "0";
%filename 'ARL_MU_2.eps';file-properties "XNPEU";}} }%
%BeginExpansion
{\parbox[b]{2.5261in}{\begin{center}
\includegraphics[
height=1.9787in,
width=2.5261in
]%
{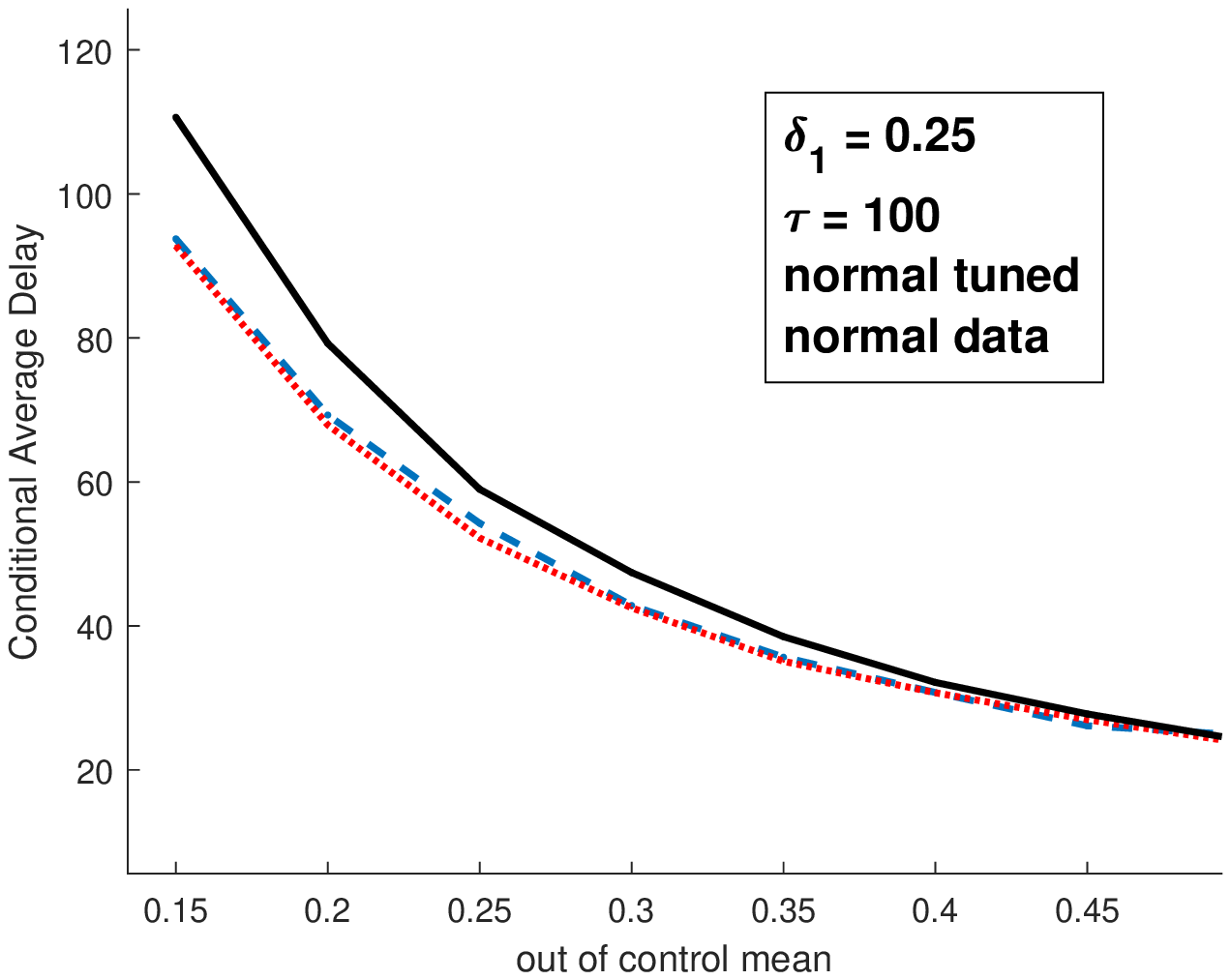}%
\\
{\protect\tiny Figure 3}%
\end{center}}}
%EndExpansion
&
%TCIMACRO{\FRAME{itbpFU}{2.6576in}{1.9501in}{0in}{\Qcb{\QTR{tiny}{Figure 4}}}%
%{}{arl_{m}u_{1}.eps}{\special{ language "Scientific Word";  type "GRAPHIC";
%display "USEDEF";  valid_file "F";  width 2.6576in;  height 1.9501in;
%depth 0in;  original-width 5.834in;  original-height 4.3708in;  cropleft "0";
%filename 'ARL_MU_1.eps';file-properties "XNPEU";}} }%
%BeginExpansion
{\parbox[b]{2.6576in}{\begin{center}
\includegraphics[
height=1.9501in,
width=2.6576in
]%
{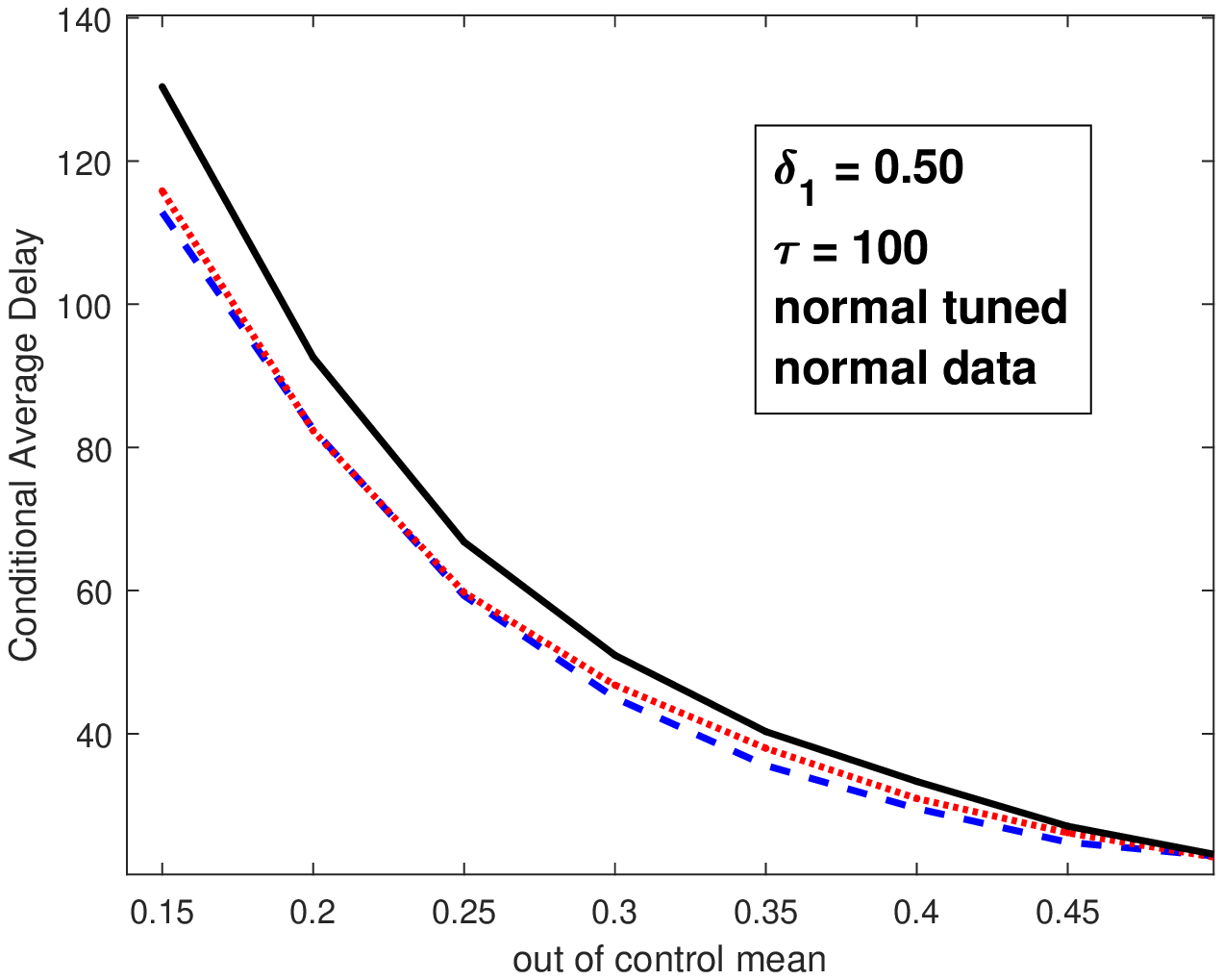}%
\\
{\protect\tiny Figure 4}%
\end{center}}}
%EndExpansion
\end{array}
$%

\begin{tabular}
[c]{ll}%
Figures 1 to 4 & Conditional average delay simulation results.\\
& Tuned for normal distribution.\\
& Data from a normal distribution.
\end{tabular}

\end{center}

The preceding discussion indicates that the differences between the CADT of
the schemes are relatively small when all three are tuned for data coming from
a normal distribution. However, it is rather interesting to see what
transpires when the data actually come from a non-normal distribution, that
is, when the schemes have been tuned to the wrong distribution. Figures 5 and
6 are the counterparts of Figures 1 and 2 for data coming from a Laplace
distribution. The most obvious difference between the two sets of figures is
that the NPSR has, in a manner of speaking, gone from "best" to "worst". All
three schemes detect the change in the mean of the data from the heavier
tailed Laplace distribution sooner than before, but the improvement in the
CADT of the S-R and CUSUM schemes is markedly better than that of the NPSR.
This finding could perhaps be paraphrased by saying that the two SSR schemes
adapt better to erroneous tuning than the NPSR.

\begin{center}
$%
\begin{array}
[c]{cc}%
%TCIMACRO{\FRAME{itbpFU}{2.3177in}{2.2771in}{0in}{\Qcb{\QTR{tiny}{Figure 5}}}%
%{}{arl_{c}hpt_{l}aplacedata_{2}.eps}{\special{ language "Scientific Word";
%type "GRAPHIC";  display "USEDEF";  valid_file "F";  width 2.3177in;
%height 2.2771in;  depth 0in;  original-width 5.834in;
%original-height 4.3708in;  cropleft "0";  croptop "1";  cropright "1";
%filename 'ARL_CHPT_LAPLACEDATA_2.eps';file-properties "XNPEU";}} }%
%BeginExpansion
{\parbox[b]{2.3177in}{\begin{center}
\includegraphics[
height=2.2771in,
width=2.3177in
]%
{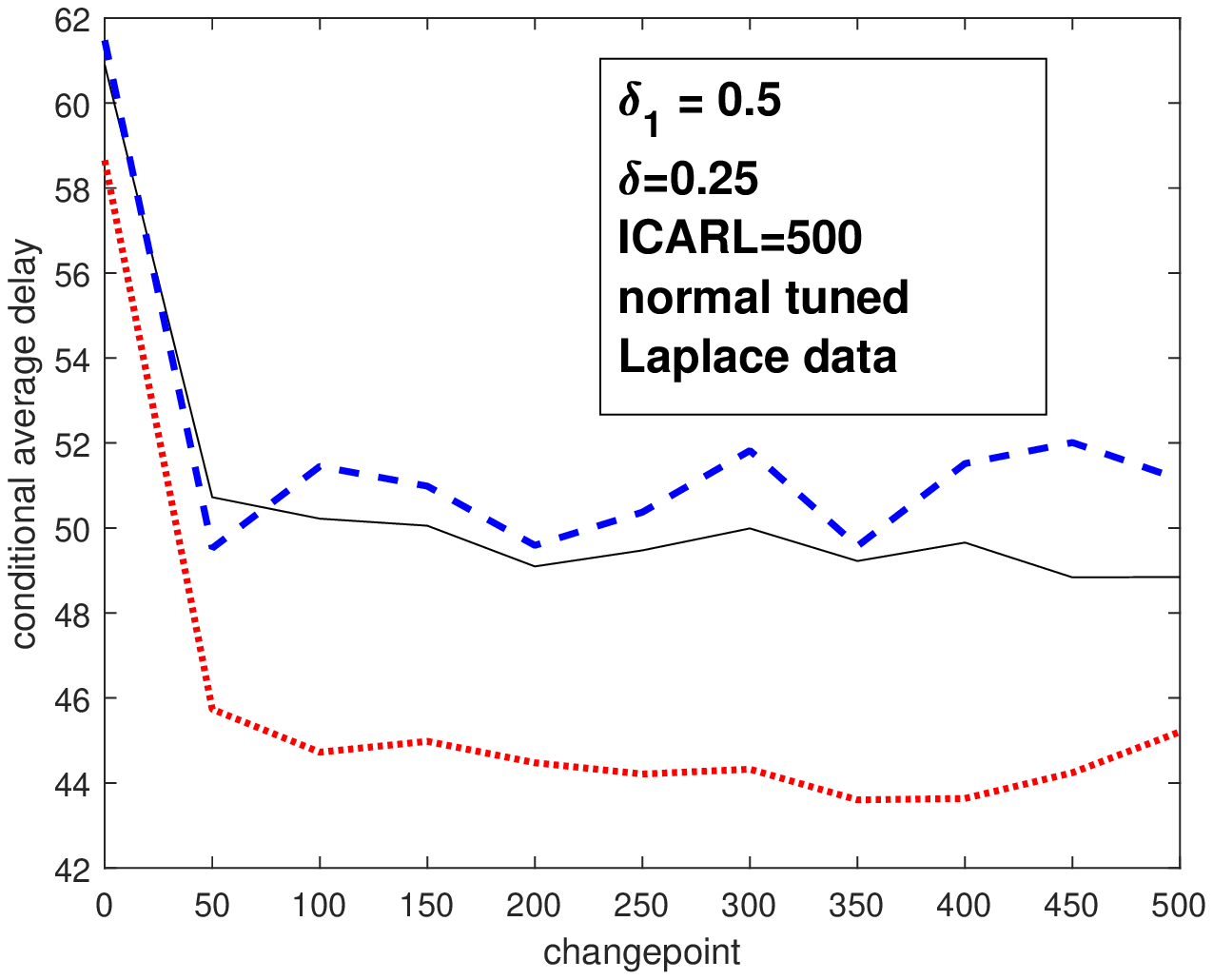}%
\\
{\protect\tiny Figure 5}%
\end{center}}}
%EndExpansion
&
%TCIMACRO{\FRAME{itbpFU}{2.4984in}{2.2355in}{0in}{\Qcb{\QTR{tiny}{Figure 6}}}%
%{}{arl_{c}hpt_{l}aplacedata_{1}.eps}{\special{ language "Scientific Word";
%type "GRAPHIC";  display "USEDEF";  valid_file "F";  width 2.4984in;
%height 2.2355in;  depth 0in;  original-width 5.834in;
%original-height 4.3708in;  cropleft "0";  croptop "1";  cropright "1";
%filename 'ARL_CHPT_LAPLACEDATA_1.eps';file-properties "XNPEU";}} }%
%BeginExpansion
{\parbox[b]{2.4984in}{\begin{center}
\includegraphics[
height=2.2355in,
width=2.4984in
]%
{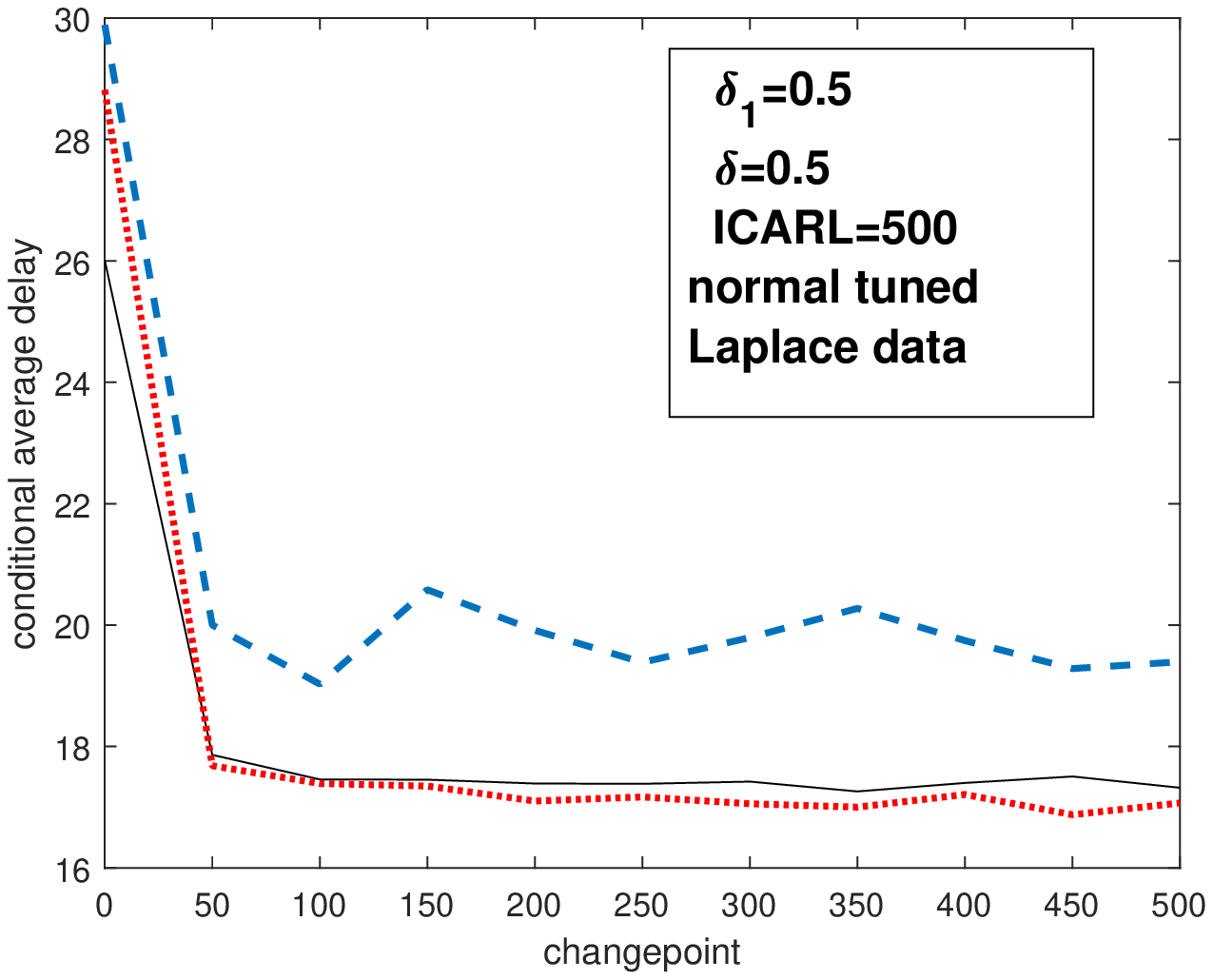}%
\\
{\protect\tiny Figure 6}%
\end{center}}}
%EndExpansion
\end{array}
$\bigskip%

\begin{tabular}
[c]{ll}%
Figures 5 and 6 & Conditional average delay simulation results.\\
& Tuned for normal distribution.\\
& Data from a Laplace distribution.
\end{tabular}

\end{center}

Next, consider a situation in which the procedures are tuned to a Laplace
distribution, for which $\theta_{0}=1.2$ in the SSR schemes. As mentioned
earlier, the use of the Laplace distribution is motivated by the NPSR scheme
which is derived from a mixture of two exponential distributions, the null
instance of which is the Laplace distribution. The tuning parameters at
$\delta_{1}=1$ are $\zeta=0.60$ for the Wilcoxon SSR S-R and CUSUM and
$(\alpha,\beta,p)=(0.57,1.00,0.88)$ for the NPSR. At $\delta_{1}=0.5$ the
parameters are $\zeta=0.3$ and $(\alpha,\beta,p)=(0.79,1.00,0.75)$. The
NPSR\ tuning parameters were obtained by exact calculation from equations (10)
and (15) in Gordon and Pollak (1994). Figures 7 and 8 show the results
when the schemes are tuned to the correct, i.e. Laplace, and incorrect, i.e.
normal, underlying distributions respectively. In Figure 7 we see that the
NPSR performs a little bit better than the S-R and a lot better than the
CUSUM, especially at early changes. When the data come from the thinner tailed
normal distribution then Figure 8 indicates, as expected, that the detection
capability of each of the three schemes degenerates considerably. However, the
S-R scheme seems to adapt itself better to the mistuning than the other two schemes.

\begin{center}
$%
\begin{array}
[c]{cc}%
%TCIMACRO{\FRAME{itbpFU}{2.2589in}{1.8741in}{0in}{\Qcb{\QTR{tiny}{Figure 7}}}%
%{}{Figure 7}{\special{ language "Scientific Word";  type "GRAPHIC";
%display "USEDEF";  valid_file "F";  width 2.2589in;  height 1.8741in;
%depth 0in;  original-width 5.834in;  original-height 4.3708in;  cropleft "0";
%filename '../GR Cusums docs/ARL_CHPT_LAPLACE_LAPLACE_2.eps';file-properties "XNPEU";}%
%} }%
%BeginExpansion
{\parbox[b]{2.2589in}{\begin{center}
\includegraphics[
height=1.8741in,
width=2.2589in
]%
{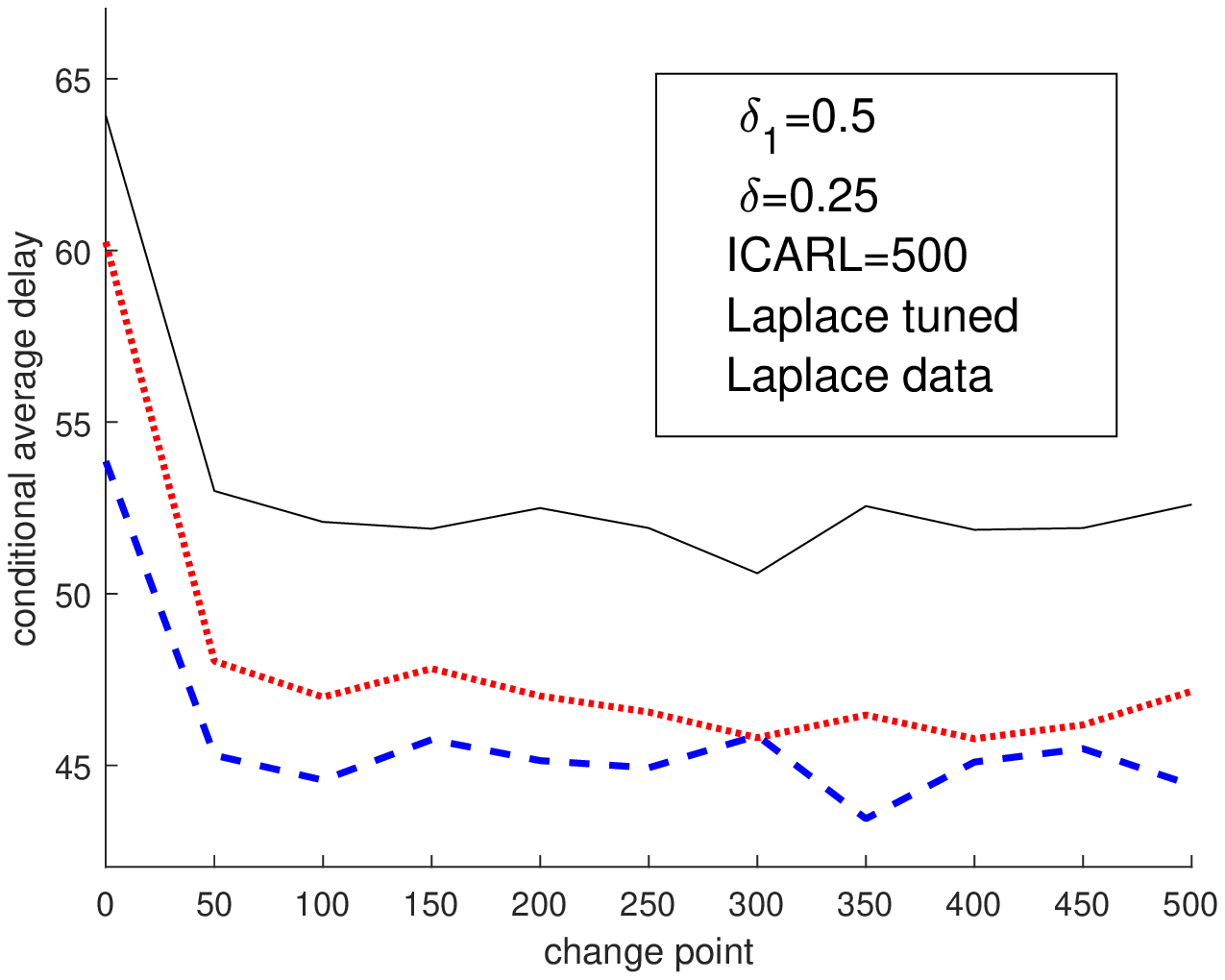}%
\\
{\protect\tiny Figure 7}%
\end{center}}}
%EndExpansion
&
%TCIMACRO{\FRAME{itbpFU}{2.2658in}{1.8948in}{0in}{\Qcb{\QTR{tiny}{Figure 8}}}%
%{}{Figure 8}{\special{ language "Scientific Word";  type "GRAPHIC";
%display "USEDEF";  valid_file "F";  width 2.2658in;  height 1.8948in;
%depth 0in;  original-width 5.834in;  original-height 4.3708in;  cropleft "0";
%filename '../GR Cusums docs/ARL_CHPT_LAPLACE_normal_2.eps';file-properties "XNPEU";}%
%} }%
%BeginExpansion
{\parbox[b]{2.2658in}{\begin{center}
\includegraphics[
height=1.8948in,
width=2.2658in
]%
{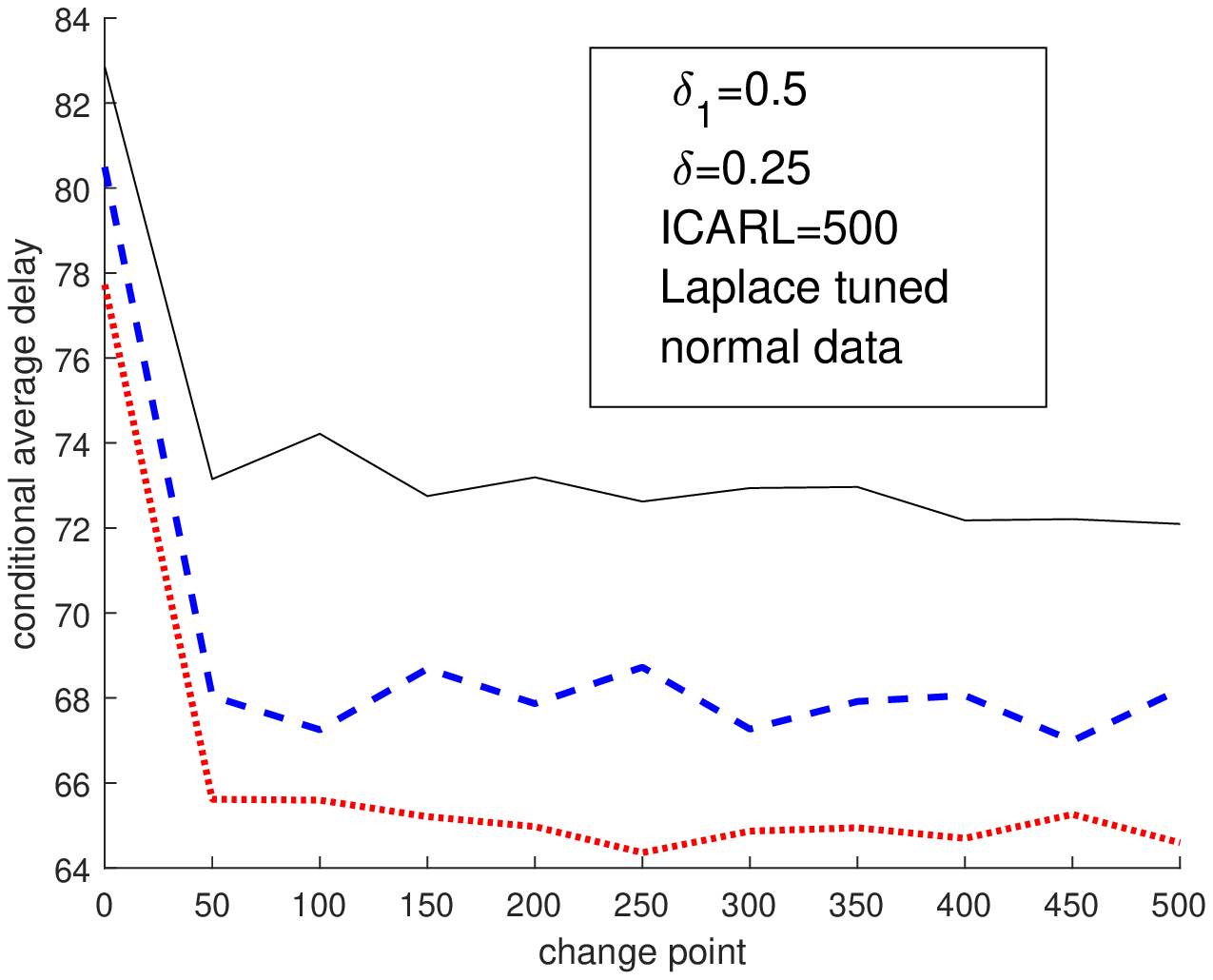}%
\\
{\protect\tiny Figure 8}%
\end{center}}}
%EndExpansion
\end{array}
$%

\begin{tabular}
[c]{ll}%
Figures 7 and 8 & Conditional average delay simulation results.\\
& Tuned for Laplace distribution.
\end{tabular}

\end{center}

Finally, we compare some stationary average delay times (SADTs). The SADT is
the average delay time measured from the time of the last false alarm,
assuming that a stationary regime has established itself. Figures 1 and 2
already suggest strongly that stationarity sets in rather quickly and that the
NPSR and SSR S-R schemes would perform similarly and exhibit smaller SADTs
than the CUSUM. Here we take $ARL_{0}=500$, a target mean $\delta_{1}=0.5$ and
change points $\tau=500,\ 1,000$ and $1,500$ which ensures many restarts,
thus\ a stationary situation, before the change takes place. For each of the
schemes, the three SADT curves corresponding to the three change points are
virtually indistinguishable, confirming what was anticipated earlier after
looking at Figures 1 and 2, namely that a form of stationarity seems to become
in force rather early. Therefore, only the results for $\tau=1,000$ are plotted here.

These results are shown in Figure 9.\ The NPSR and S-R schemes are seen to
behave very similarly and to have CADTs that are substantially smaller than
those of the SSR CUSUM at shifts that are substantially less than the target.
That the SADT performance of the S-R scheme is better than that of the CUSUM
is a result that is in line with what is known about the behaviour of the
corresponding parametric schemes - see Moustakides, Polunchenko and
Tartakovsky (2009). Finally, Figure 10 shows that if the data actually come
from a Laplace, rather than a normal, distribution then the NPSR again goes
from "best" to "worst".

\begin{center}
$%
\begin{array}
[c]{cc}%
%TCIMACRO{\FRAME{itbpFU}{2.1369in}{1.6864in}{0in}{\Qcb{\QTR{tiny}{Figure 9}}}%
%{}{sadt_{n}ormal_{n}ormal_{p}oint5.eps}{\special{ language "Scientific Word";
%type "GRAPHIC";  display "USEDEF";  valid_file "F";  width 2.1369in;
%height 1.6864in;  depth 0in;  original-width 5.834in;
%original-height 4.3708in;  cropleft "0";  croptop "1";  cropright "1";
%filename 'SADT_normal_normal_point5.eps';file-properties "XNPEU";}} }%
%BeginExpansion
{\parbox[b]{2.1369in}{\begin{center}
\includegraphics[
height=1.6864in,
width=2.1369in
]%
{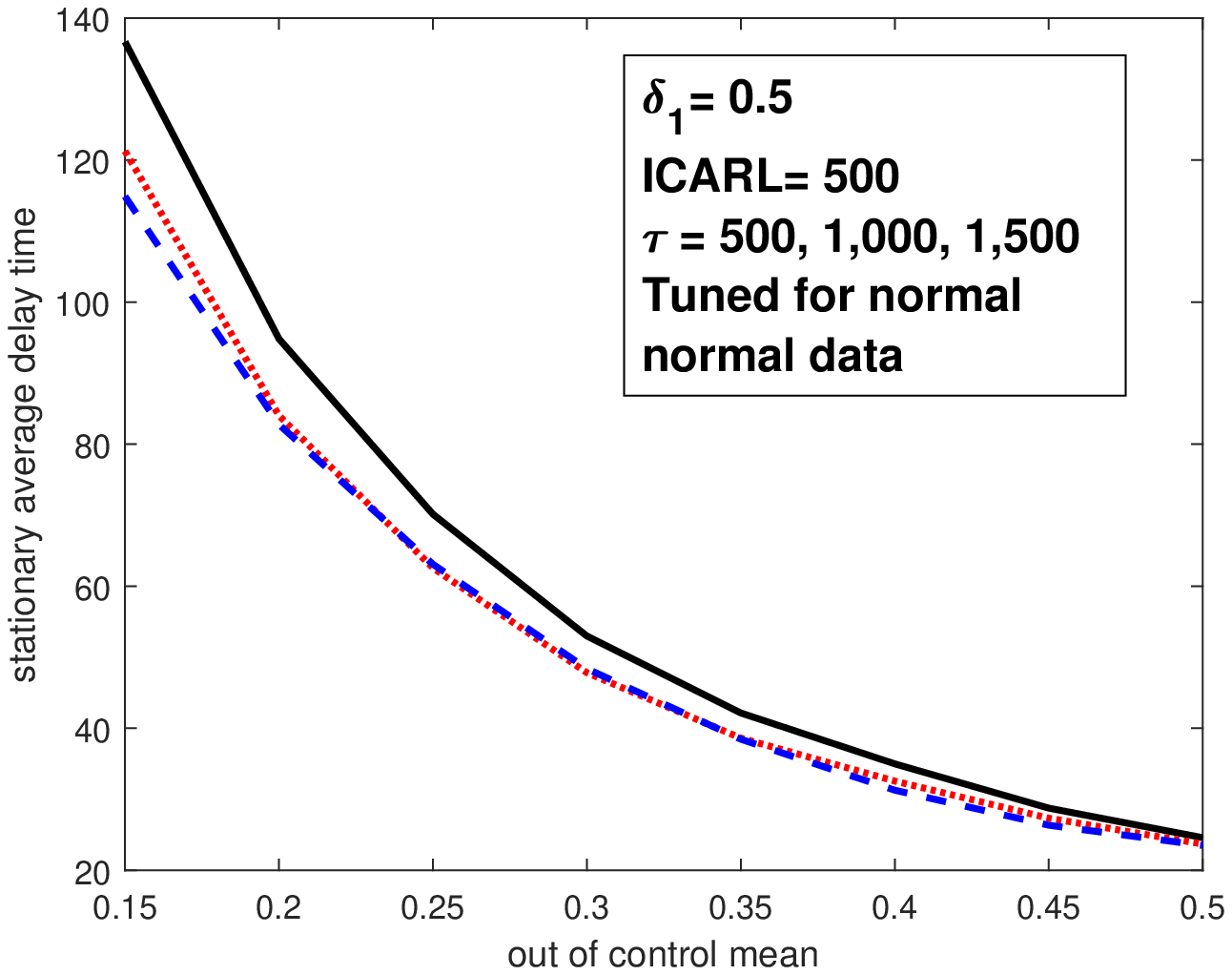}%
\\
{\protect\tiny Figure 9}%
\end{center}}}
%EndExpansion
&
%TCIMACRO{\FRAME{itbpFU}{2.13in}{1.6656in}{0in}{\Qcb{\QTR{tiny}{Figure 10}}}%
%{}{sadt_{n}ormal_{l}aplace_{p}oint5.eps}%
%{\special{ language "Scientific Word";  type "GRAPHIC";  display "USEDEF";
%valid_file "F";  width 2.13in;  height 1.6656in;  depth 0in;
%original-width 5.834in;  original-height 4.3708in;  cropleft "0";
%filename 'SADT_normal_Laplace_point5.eps';file-properties "XNPEU";}} }%
%BeginExpansion
{\parbox[b]{2.13in}{\begin{center}
\includegraphics[
height=1.6656in,
width=2.13in
]%
{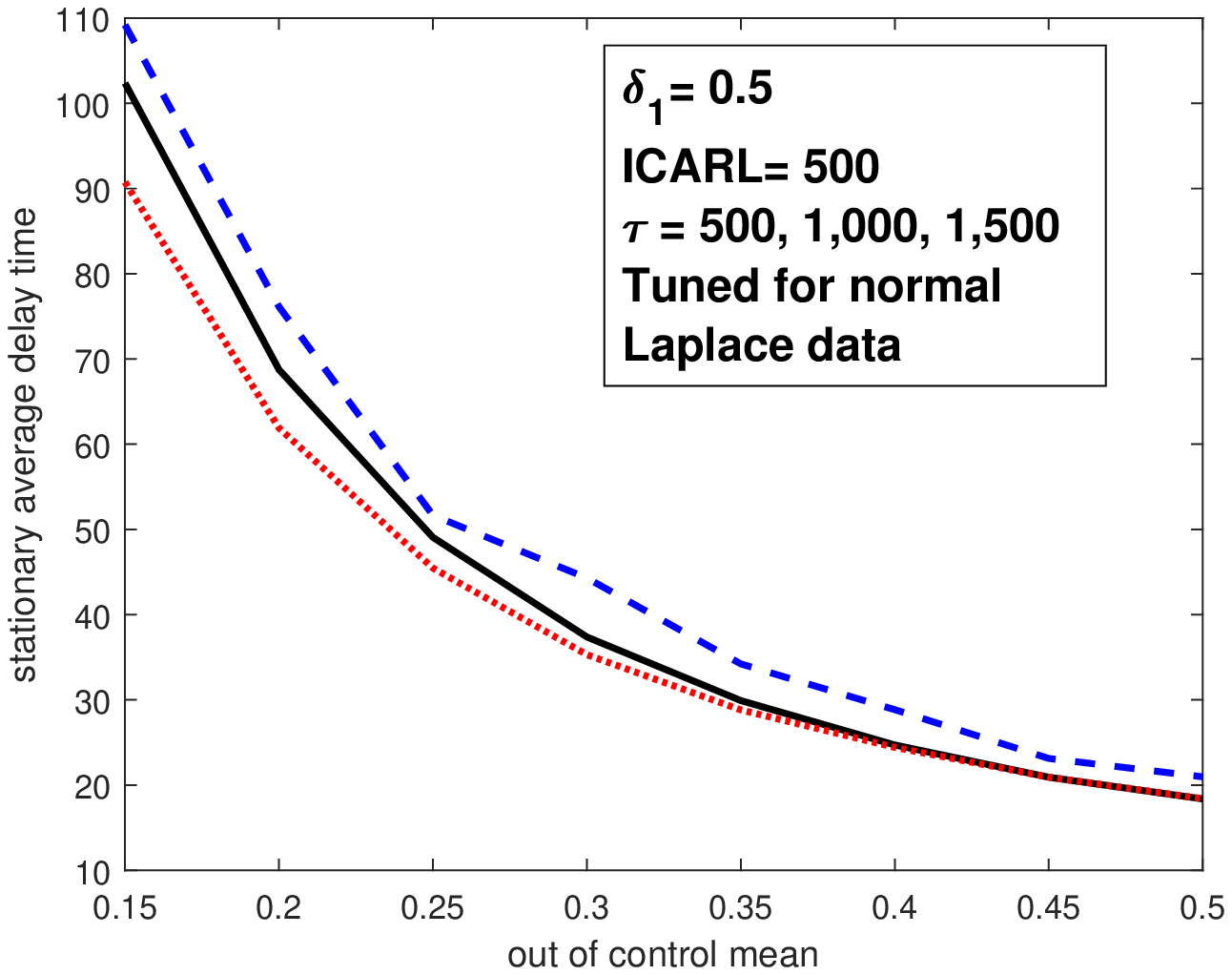}%
\\
{\protect\tiny Figure 10}%
\end{center}}}
%EndExpansion
\end{array}
$%

\begin{tabular}
[c]{ll}%
Figures 9 and 10 & Stationary average delay time simulation results.
\end{tabular}

\end{center}

\section{Application}

The data and the particular application from which they arose are similar to
those examined by Lombard and Van Zyl (2018) and consist of sequentially
observed pairs of replicate coal ash values $(V_{1i},V_{2i}),\ i\geq1$. The
measurements $V_{1}$ and $V_{2}$ come from two nominally identical coal
samples analyzed by two independent laboratories. Denote the observations from
the two laboratories by $V_{k}=T+\epsilon_{k},\ k=1,\ 2$ where $T$ is the true
ash content and the $\epsilon_{k}$ denote measurement error. These errors
should be independent and identically distributed with zero means and common
standard deviation $\sigma$. Then the difference $X=V_{1}-V_{2}=\epsilon
_{1}-\epsilon_{2}$ is independent of $T$ and should be symmetrically
distributed around zero. However, if the mean of $X$ is nonzero then there
exists a bias between the laboratory results which would call for an audit of
their respective methodologies to identify the cause of the bias. Therefore,
our interest is in monitoring the observed $X_{i}$ sequence for sustained
deviations away from a zero mean. The preceding is a typical matched pairs
setup in which, were we dealing with a fixed sample of pairs, the Wilcoxon
signed rank test would typically be used. Since the data are accruing one pair
at a time, use of some sequential version of the test seems appropriate. We
will compare the results produced by the Wilcoxon SSR S-R, SSR CUSUM and NPSR.

The three schemes considered thus far were designed to detect positive shifts.
Denote by $S^{(k)}(X),\ k=1,2,3$, any one of these schemes applied to the data
$X=(X_{1},X_{2},\ldots)$. To also detect negative shifts, we run
simultaneously the schemes $S^{(k)}(X)$ and $S^{(k)}(-X)$, that is, two
schemes with the same $\zeta$ and $h$, one on the $X$ data and the other on
the sign-changed data $-X$. This is then the two-sided scheme with run length
the smaller of the two constituent run lengths. The resulting ICARL is often
close to one half that of each of the one-sided schemes. Thus, to find the
control limits that produce a nominal $ARL_{0}$ in a two-sided scheme, the
control limit applicable to a nominal $2\times ARL_{0}$ in a one-sided scheme,
adjusted after some Monte Carlo simulation, is used.

A practical limitation of the NPSR is that it is not computationally feasible
to generate sufficiently accurate control limits guaranteeing ICARLs of $1,000$
or more at a wide range of reference values $\zeta$. The data set shown in
Figure 11, indicates \textit{in retrospect} a change point relatively soon after
initialization. Thus, we will implement the S-R, the CUSUM and the NPSR, using a
two-sided $ARL_{0}$ of $400$, making implementation of the NPSR feasible. In
all three schemes the target change size is set at $\delta_{1}=0.25$ and each
scheme is tuned to a $t_{4}$ distribution. For the S-R and CUSUM, this results
in a reference value $\zeta=0.15$ (rounded to two decimal places) and control
limits $h=6.52$ for the S-R and $ h=11.3$ for the CUSUM, found by
interpolation from Tables 4 and 5. The tuning constants for the NPSR at $\delta_{1}=0.25$, namely $\alpha
=0.8912$, $\beta=1.0774$ and $p=0.6292$, were found by numerical computation
from formulas (10) and (15) in Gordon and Pollak (1994). The approximation
\[
\alpha\times ARL_{0}=0.8912\times800=712.96
\]
to the ICARL of the (one-sided) NPSR (Gordon and Pollak, 1994, Theorem 2.2),
supplemented with some Monte Carlo simulation, leads to a control limit
$h=725$ for an $ARL_{0}=400$ in a two-sided NPSR.%

%TCIMACRO{\FRAME{dtbpFU}{3.5743in}{2.4154in}{0pt}{\Qcb{Figure 11 \ Plot of
%$150$ successive pairwise ash differences.}}{}{ashdata.eps}%
%{\special{ language "Scientific Word";  type "GRAPHIC";  display "USEDEF";
%valid_file "F";  width 3.5743in;  height 2.4154in;  depth 0pt;
%original-width 5.834in;  original-height 4.3708in;  cropleft "0";
%filename '../GR Cusums docs/Ashdata.eps';file-properties "XNPEU";}} }%
%BeginExpansion
\begin{center}
\includegraphics[
height=2.4154in,
width=3.5743in
]%
{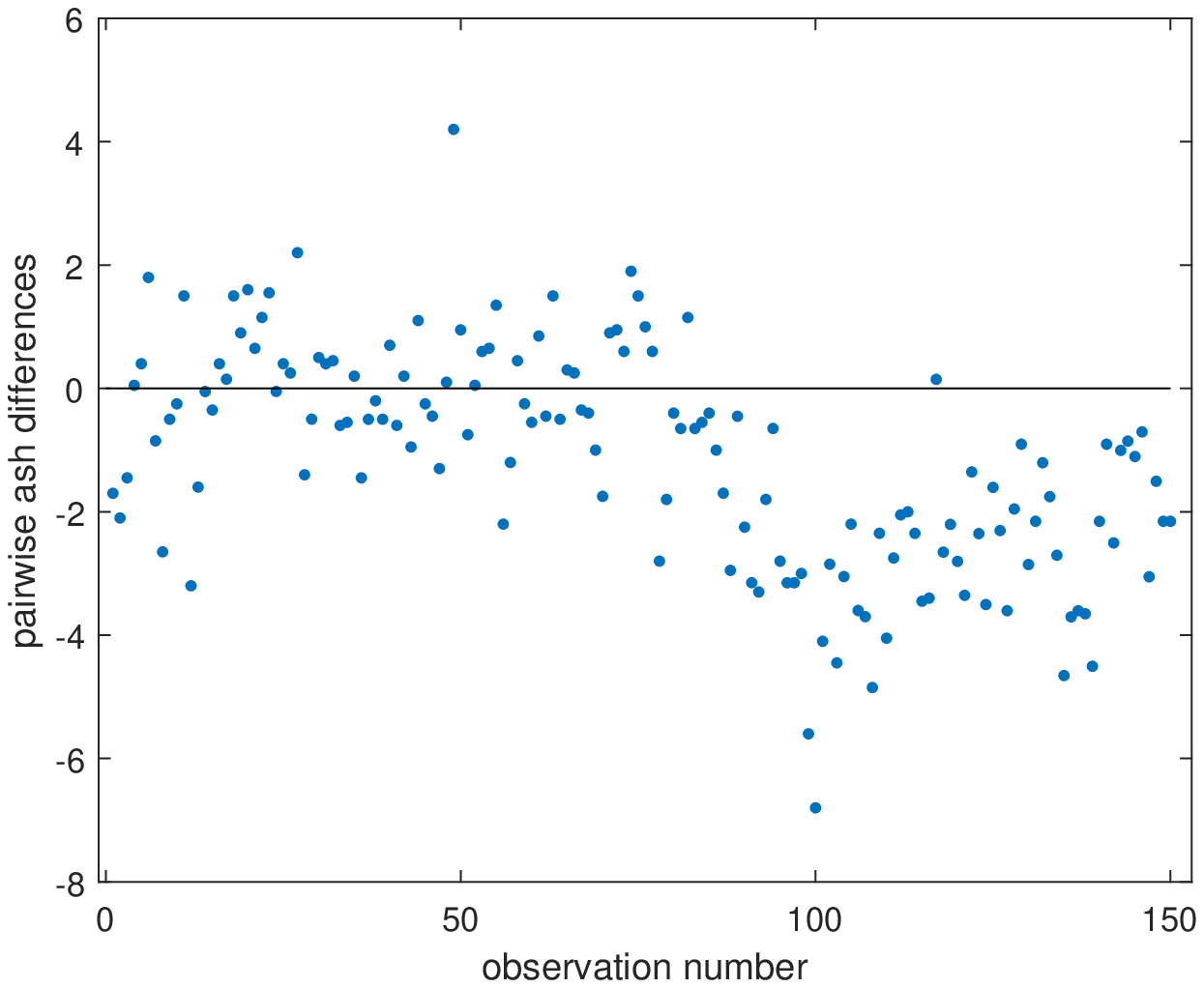}%
\\
Figure 11 \ Plot of $150$ successive pairwise ash differences.
\end{center}
%EndExpansion

Plots of the paths of the three schemes, applied to the data shown in Figure
11 are in Figures 12, 13 and 14. For visual presentation we plot $S^{(k)}(X)$
with control limit $h$ (the upper path) and $-S^{(k)}(-X),\ i\geq1$ with
control limit $-h$ (the lower path). The CUSUM scheme sounds an alarm at
$i=91$ while the S-R and NPSR both do so at $i=93$. These conclusions seem to
be in agreement with what is seen in Figure 11.

A distinguishing feature in a two-sided normal distribution CUSUM, also
evident in Figure 12, is that the upper (lower) CUSUM is at zero whenever the
lower (upper) CUSUM is non-zero. The usual CUSUM change point estimator is then
the last index at which the hitting CUSUM sequence, upper or lower, was at
zero. In the present instance, this estimator gives $\tau=77$ as the
change point. The same feature is not present in the S-R or the NPSR, so that
an alternative estimator must be sought. A straightforward approach is to look
upon the stopped sequence $X_{1},X_{2},\ldots,X_{N}$ as consisting of samples
$\{X_{1},\ldots,X_{\tau}\}$ and $\{X_{\tau+1},\ldots,X_{N}\}$ from two
distributions differing only in location and to estimate $\tau$ by least
squares, conveniently ignoring the fact that the observed run length $N$ is,
in fact, a random variable. Then the least squares estimator of $\tau$ is
\begin{equation}
\widehat{\tau}=arg\ max_{1\leq k\leq N-1}\ |T_{k}| \label{LS tau est}%
\end{equation}
where
\begin{equation}
T_{k}=\sum_{i=k+1}^{N}X_{i}/\sqrt{N-k}. \label{Tk}%
\end{equation}
This suggests using the estimator (\ref{LS tau est}) after replacing $X_{i}$
in (\ref{Tk}) by $\xi_{i}$ from (\ref{SSRxi}). Denote this version of $T_{k}$
by $T_{k}^{\ast}$. A plot of $|T_{k}^{\ast}|$ against $k$ is shown in Figure
15. The maximum occurs at $\widehat{\tau}=78$, which is almost the same
estimate as that found from the CUSUM.%

%TCIMACRO{\FRAME{dtbpFU}{3.179in}{2.687in}{0pt}{\Qcb{Figure 12 \ Wilcoxon SSR
%CUSUM paths.}}{}{wilcoxon_ssr_cusum_2.eps}%
%{\special{ language "Scientific Word";  type "GRAPHIC";  display "USEDEF";
%valid_file "F";  width 3.179in;  height 2.687in;  depth 0pt;
%original-width 5.834in;  original-height 4.3708in;  cropleft "0";
%filename '../GR Cusums docs/Wilcoxon_SSR_CUSUM_2.eps';file-properties "XNPEU";}%
%} }%
%BeginExpansion
\begin{center}
\includegraphics[
height=2.687in,
width=3.179in
]%
{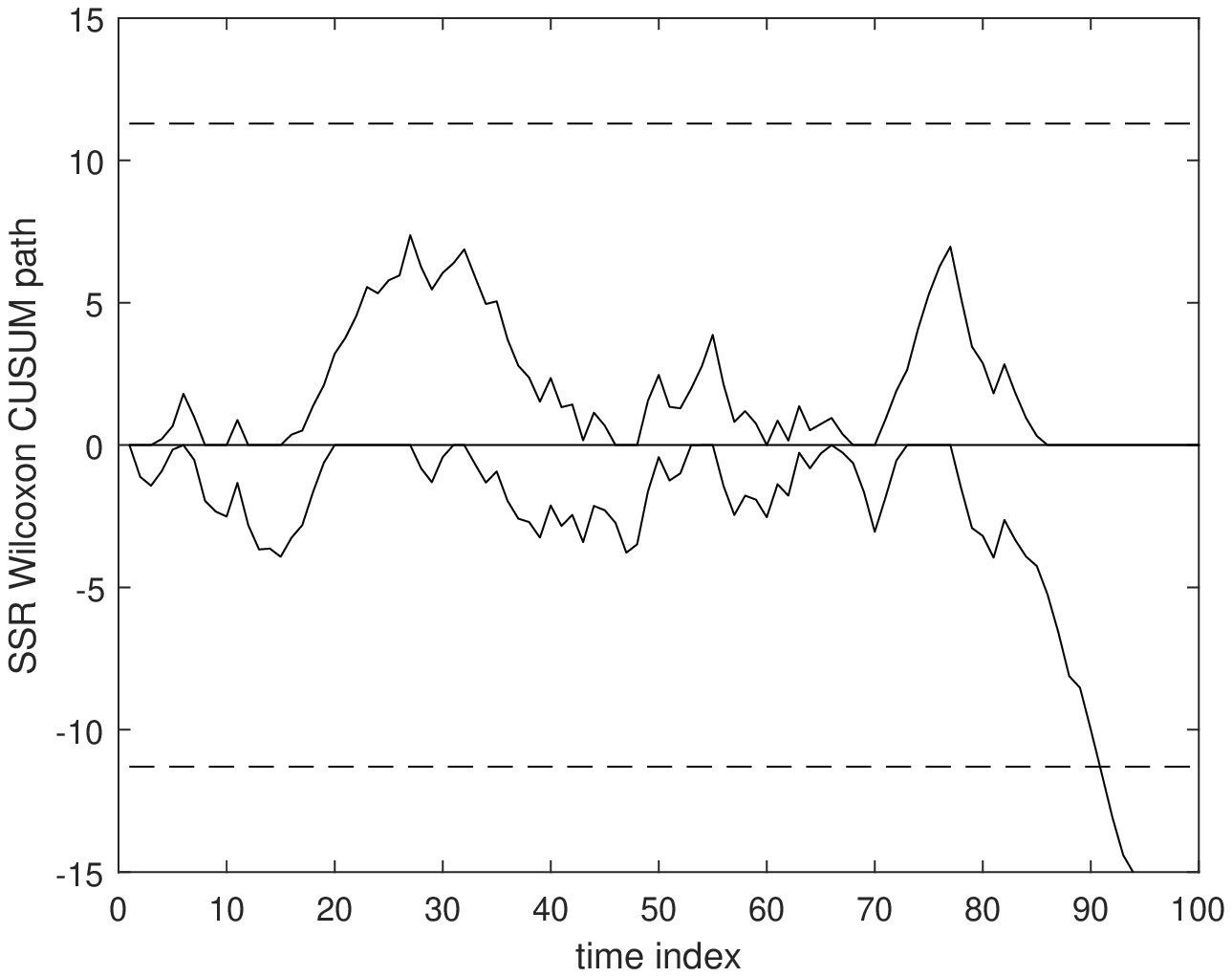}%
\\
Figure 12 \ Wilcoxon SSR CUSUM paths.
\end{center}
%EndExpansion
%TCIMACRO{\FRAME{dtbpFU}{3.1721in}{2.6446in}{0pt}{\Qcb{Figure 13 \ Wilcoxon SSR
%S-R paths.}}{}{wilcoxon_ssr_s-r_2.eps}{\special{ language "Scientific Word";
%type "GRAPHIC";  display "USEDEF";  valid_file "F";  width 3.1721in;
%height 2.6446in;  depth 0pt;  original-width 5.834in;
%original-height 4.3708in;  cropleft "0";  croptop "1";  cropright "1";
%filename '../GR Cusums docs/Wilcoxon_SSR_S-R_2.eps';file-properties "XNPEU";}}
%}%
%BeginExpansion
\begin{center}
\includegraphics[
height=2.6446in,
width=3.1721in
]%
{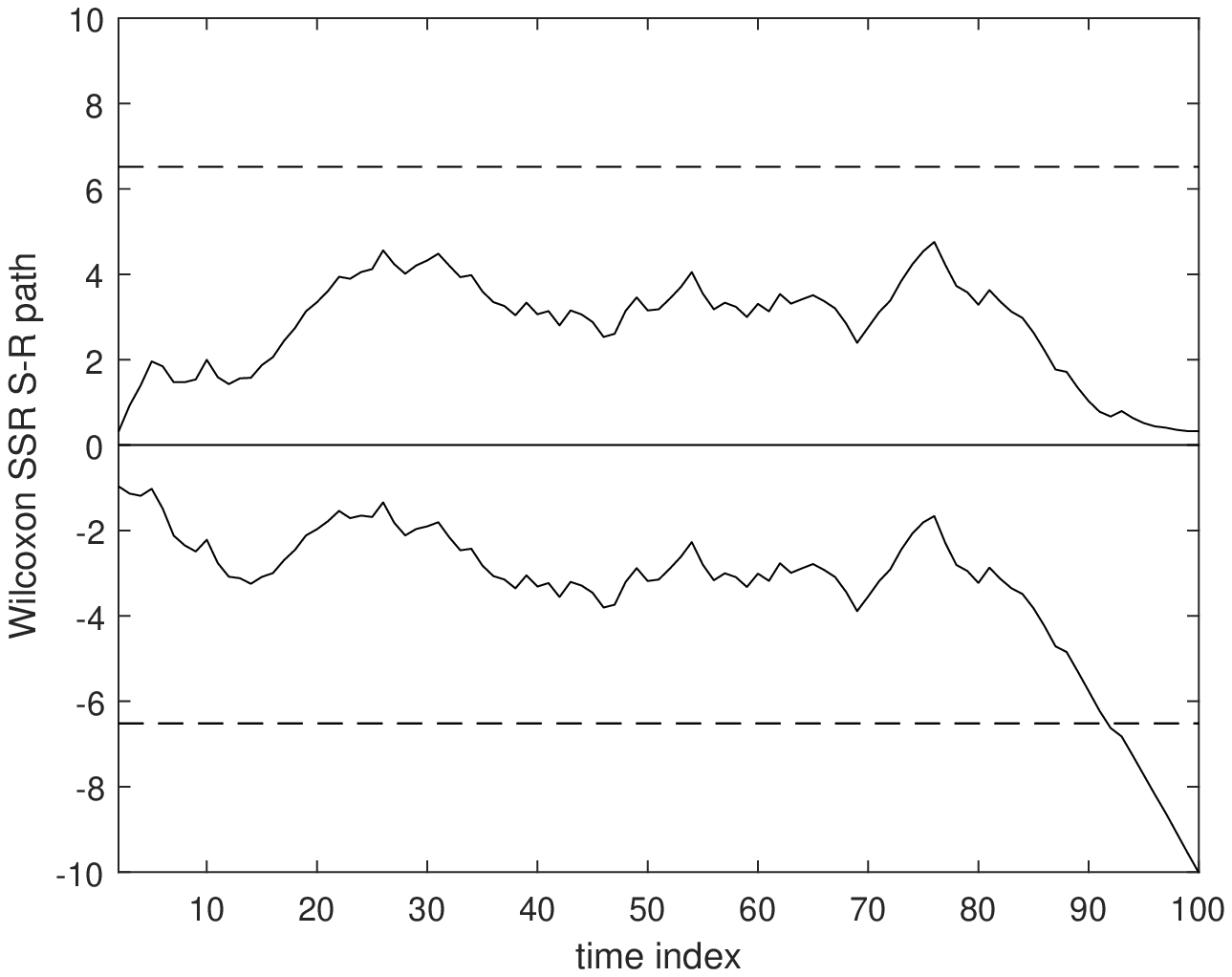}%
\\
Figure 13 \ Wilcoxon SSR S-R paths.
\end{center}
%EndExpansion
%

%TCIMACRO{\FRAME{dtbpFU}{3.269in}{2.2978in}{0pt}{\Qcb{Figure 14 \ \ NPSR paths.
%\ }}{}{gp_2.eps}{\special{ language "Scientific Word";  type "GRAPHIC";
%display "USEDEF";  valid_file "F";  width 3.269in;  height 2.2978in;
%depth 0pt;  original-width 5.834in;  original-height 4.3708in;  cropleft "0";
%filename '../GR Cusums docs/GP_2.eps';file-properties "XNPEU";}} }%
%BeginExpansion
\begin{center}
\includegraphics[
height=2.2978in,
width=3.269in
]%
{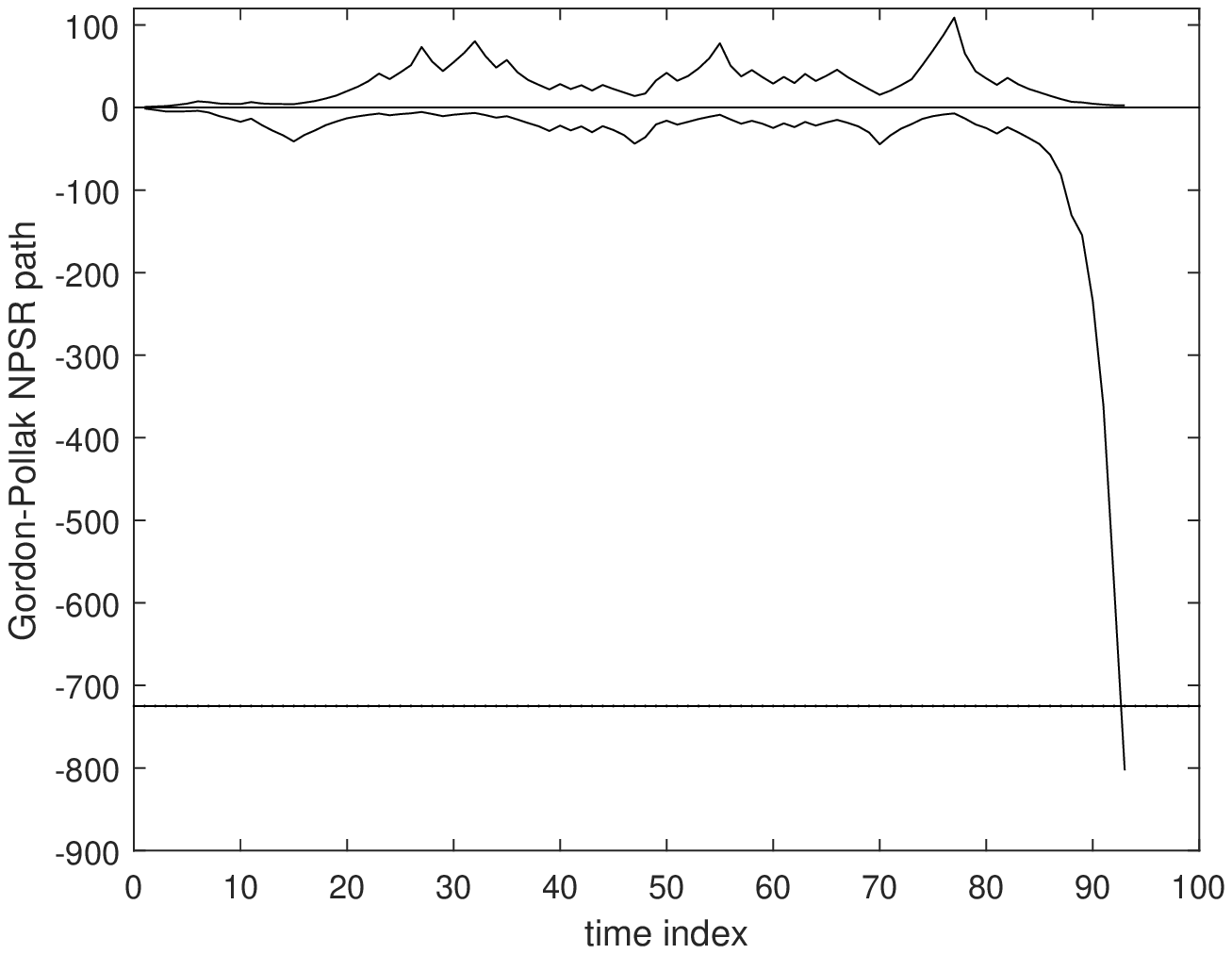}%
\\
Figure 14 \ \ NPSR paths. \
\end{center}
%EndExpansion
%

%TCIMACRO{\FRAME{dtbpFU}{3.1341in}{2.2191in}{0pt}{\Qcb{Figure\ 15 \ Least
%squares estimate of the change point.}}{}{tau_est_2.eps}%
%{\special{ language "Scientific Word";  type "GRAPHIC";  display "USEDEF";
%valid_file "F";  width 3.1341in;  height 2.2191in;  depth 0pt;
%original-width 5.834in;  original-height 4.3708in;  cropleft "0";
%filename '../GR Cusums docs/tau_est_2.eps';file-properties "XNPEU";}} }%
%BeginExpansion
\begin{center}
\includegraphics[
height=2.2191in,
width=3.1341in
]%
{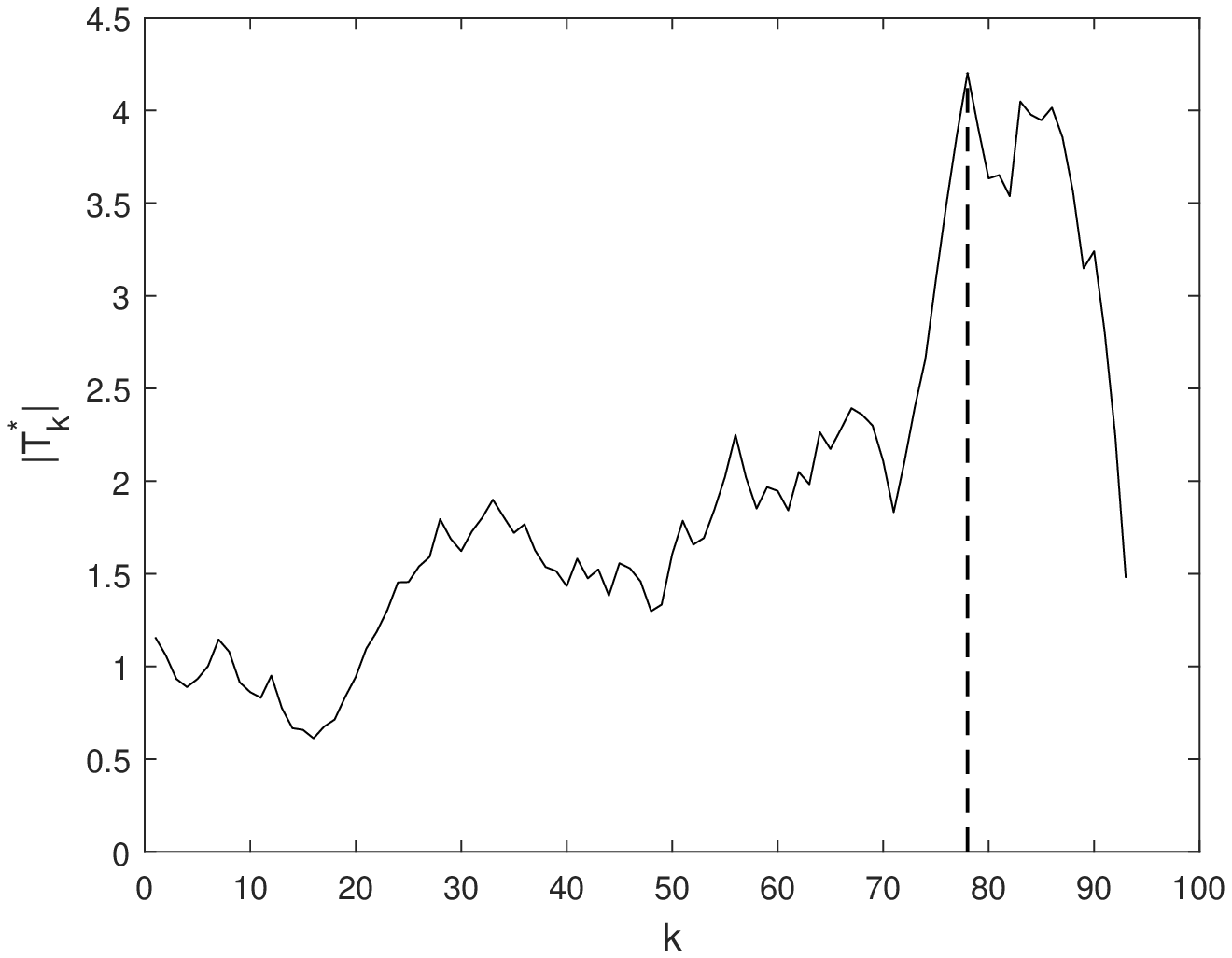}%
\\
Figure\ 15 \ Least squares estimate of the change point.
\end{center}
%EndExpansion

\section{Summary}

We develop a Shiryaev-Roberts type scheme based on signed sequential ranks for
detecting a change in the median from zero to a nonzero value in an
unspecified symmetric distribution. The scheme is distribution free and
scale invariant, meaning that a single set of control limits apply regardless
of the functional form of the underlying distribution. Monte Carlo simulation
results indicate that the scheme performs very well under a broad range of
circumstances. In particular, it seems to be more adept at detecting small
changes than a corresponding signed sequential rank CUSUM. Some Monte Carlo
simulations involving the signed sequential rank schemes and the distribution-free 
NPSR scheme developed by Gordon and Pollak (1994) suggest that the
distribution-free Shiryaev-Roberts and NPSR schemes often show better
performance in terms of out-of-control run length than the distribution-free
CUSUM. The conceptual and computational simplicity of the distribution-free
Shiryaev-Roberts scheme makes it an attractive alternative to both the distribution-free CUSUM 
and NPSR schemes. The focus in this paper has been on detecting a location change in a
symmetric distribution. A matter for further research is the possibility of
detecting scale changes via an appropriate construction of a Shiryaev-
Roberts type sequential rank scheme. Furthermore, the possibility of
constructing such schemes to deal with the detection of location and scale
changes in asymmetric distributions, needs to be investigated.

\newpage

\section{Appendix}

{\large Derivation of eqn. (\ref{OOCmean_ksi})}\bigskip

It is convenient to let $Y,Y_{1},Y_{2},\ldots$ be i.i.d. with a common
symmetric around zero distribution with cdf $F$. Then, if a shift of size
$\delta\neq0$ occurs at index $\tau+1$, the observations can be represented as
$X_{1}=Y_{1},\ldots,X_{\tau}=Y_{\tau},$ $X_{\tau+1}=Y_{\tau+1}+\delta$.
Furthermore, then%
\begin{align*}
&  E\left[  s(Y_{\tau+1}+\delta)\boldsymbol{1}\left(  |Y_{j}|<|Y_{\tau
+1}+\delta|\right)  |Y_{\tau+1}=y\right] \\
&  =s(y+\delta)\Pr\left(  |Y|<|y+\delta|\right) \\
&  =s(y+\delta)\left(  2F(|y+\delta|)-1\right) \\
&  =2F(y+\delta)-1.
\end{align*}
Consequently,%
\begin{align}
E\left[  \nu_{\tau+1}\xi_{\tau+1}\right]   &  =\frac{1}{\tau+2}E\left[
s(X_{\tau+1})%
%TCIMACRO{\tsum \nolimits_{j=1}^{\tau}}%
%BeginExpansion
{\textstyle\sum\nolimits_{j=1}^{\tau}}
%EndExpansion
\boldsymbol{1}\left(  |Y_{j}|\leq|Y_{\tau+1}+\delta|\right)  \right]
\nonumber\\
&  =\frac{\tau}{\tau+2}E\left[  2F(Y+\delta)-1\right]  +\frac{E\left[
s(X_{\tau+1})\right]  }{\tau+2}\nonumber\\
&  =E\left[  2F(Y+\delta)-1\right]  +0\left(  \frac{1}{\tau}\right)
\label{A1}%
\end{align}
and by Taylor expansion%
\begin{align}
E\left[  2F(Y+\delta)-1\right]   &  =E\left[  2F(Y)-1\right]  +2\delta
E\left[  f(Y)\right]  +0\left(  \delta^{2}\right) \nonumber\\
&  =2\delta%
%TCIMACRO{\tint \nolimits_{-\infty}^{+\infty}}%
%BeginExpansion
{\textstyle\int\nolimits_{-\infty}^{+\infty}}
%EndExpansion
f^{2}(y)dy+0\left(  \delta^{2}\right)  . \label{A2}%
\end{align}
Also,%
\begin{equation}
\nu_{\tau}=\sqrt{3}+0\left(  \frac{1}{\tau}\right)  . \label{A3}%
\end{equation}

Putting (\ref{A1}), (\ref{A2}) and (\ref{A3}) together, we get%
\[
E\left[  \xi_{\tau+1}\right]  =\sqrt{12}\delta%
%TCIMACRO{\tint \nolimits_{-\infty}^{+\infty}}%
%BeginExpansion
{\textstyle\int\nolimits_{-\infty}^{+\infty}}
%EndExpansion
f^{2}(y)dy+0\left(  \delta^{2}\right)  +0\left(  \frac{1}{\tau}\right)  .
\]

\end{document}